\documentclass[aps, prd, twocolumn, superscriptaddress, nofootinbib]{revtex4}
\usepackage{graphicx}% Include figure files
\usepackage{dcolumn}% Align table columns on decimal point
\usepackage{amssymb}% outlined letters
\usepackage{amsmath}% outlined letters
\usepackage{array}
\usepackage{float}

\begin{document}

%Macros
\newcommand{\Eq}[1]{\mbox{Eq. (\ref{eqn:#1})}}
\newcommand{\Fig}[1]{\mbox{Fig. \ref{fig:#1}}}
\newcommand{\Sec}[1]{\mbox{Sec. \ref{sec:#1}}}

\newcommand{\PHI}{\phi}
\newcommand{\PhiN}{\Phi^{\mathrm{N}}}
\newcommand{\vect}[1]{\mathbf{#1}}
\newcommand{\Del}{\nabla}
\newcommand{\unit}[1]{\;\mathrm{#1}}
\newcommand{\x}{\vect{x}}
\newcommand{\ScS}{\scriptstyle}
\newcommand{\ScScS}{\scriptscriptstyle}
\newcommand{\xplus}[1]{\vect{x}\!\ScScS{+}\!\ScS\vect{#1}}
\newcommand{\xminus}[1]{\vect{x}\!\ScScS{-}\!\ScS\vect{#1}}
\newcommand{\diff}{\mathrm{d}}

\newcommand{\be}{\begin{equation}}
\newcommand{\ee}{\end{equation}}
\newcommand{\bea}{\begin{eqnarray}}
\newcommand{\eea}{\end{eqnarray}}
\newcommand{\vu}{{\mathbf u}}
\newcommand{\ve}{{\mathbf e}}

        \newcommand{\vU}{{\mathbf U}}
        \newcommand{\vN}{{\mathbf N}}
        \newcommand{\vB}{{\mathbf B}}
        \newcommand{\vF}{{\mathbf F}}
        \newcommand{\vD}{{\mathbf D}}
        \newcommand{\vg}{{\mathbf g}}
        \newcommand{\va}{{\mathbf a}}

%=====================================================================
%=====================================================================
%=====================================================================

\title{Parameterised free functions and saddle stresses in modified gravity}
%Opening up the Parameter Space of MOND}
%An Exploration of Different Free Functions in MOND}
%Testing MOND with different $\mu$ functions}

\newcommand{\addressImperial}{Theoretical Physics, Blackett Laboratory, Imperial College, London, SW7 2BZ, United Kingdom}

\author{Ali Mozaffari}
\email{ali.mozaffari05@ic.ac.uk}
\affiliation{\addressImperial}

\date{\today}

\begin{abstract}
Building on previous work, we explore the parameter space of free functions in non-relativistic modified gravity theories more widely, showing that in fact the two broad regimes present have similar functional forms between different models.  Using different parameterisations, we investigate the effects on scaling tidal stresses as well as attempt to constrain the (hitherto poorly understood) deep MONDian scaling $C$.  We also consider a new intermediate MOND limit in these theories and what it tells us about the transition between these regimes. Finally we suggest a model independent framework, with the aim of constraining the MONDian parameter space using future data, such as the forthcoming LISA Pathfinder mission.  
 
\end{abstract}


\maketitle

%=====================================================================
%=====================================================================
%=====================================================================

\section{Introduction}
The concordance model of modern cosmology rests soundly on two cornerstones, a universe filled mostly with cold dark matter and dark energy (described
by a cosmological constant), ie $\Lambda$CDM, with underlying dynamics characterised by Einstein's theory of General Relativity (GR).  While this model explains the early universe with ever increasing accuracy, as long as there remains the lack of direct detection of a dark matter particle (baring unviable candidates such as neutrinos~\cite{DMastro}), it remains prudent to consider alternatives.  One such pathway available is to modify the underlying dynamics themselves, subject to the condition that above certain scales we restore our familiar Newtonian limit.  MOdified Newtonian Dynamics~\cite{Milgrom:1983ca,aqual} (MOND) provides just such a scheme.  In the past decade, the potential accomplishments of MONDian theories have been put on a pedestal equal to GR with the development of fully relativistic modified gravity theories, examples of such include Bekenstein's ground breaking theory of T$e$V$e$S~\cite{teves}, Einstein \AE
ther theories~\cite{jacobmatAE, aether} and their generalised friends~\cite{aether1,aether2,AESS} as well as bimetric theories~\cite{bimetric} and various others~\cite{BSTV,Clifton11,Fam-gaugh}.  While the MONDian paradigm provides a useful framework for making connection to observables, the free functions and parameters in these theories remain relatively unconstrained, leading to a %(not unfamiliar to cosmologists)
problem of fine-tuning.  Much work has been done investigating MONDian effects
on the largest scales, for instance applying constraints from galactic data when seeking dark matter alternatives~\cite{Zhao,binney,yusaf,yusaf1,yusaf2},
the much hailed bullet cluster has been considered for what it can tell us about the necessity or needlessness of dark matter and MOND~\cite{Angus,bullet,bullet1,bullet2,bullet3},
as has applying Lorentz violating mechanisms (typically well constrained in the matter sector) to the gravity sector~\cite{withers, lviolationcosmo}.
Deviations from the inverse square law~\cite{Blanchet, Sereno,Milgromss} have also been considered; however, little more seems to be known about constraining modified gravity theories purely in the solar system.  A chance of extending the forthcoming LISA Pathfinder (LPF) mission~\cite{LISA,companion}, to include probing the low acceleration regime around gravitational saddle points, provides just such an opportunity, both for testing and also cleanly constraining these theories.

Previous work has explored the potential of using LPF measurements to test MONDian theories~\cite{bevis,bekmag}.  We investigated the expected signal-to-noise ratio (SNR) from LPF for such a test, as well as considering effects from a variety of instrumental noise models, trajectories past the saddle and issues arising from systematics such as self-gravity~\cite{ali}.  We also considered the possibility of constraining MOND in the event of a null result, as well as rescaling parameters in theories with a preferred acceleration scale~\cite{MagAliscaling}.  By investigating the symmetries present in such a scheme, we developed a simple algorithm for scaling tidal stresses in the event of parameters $a_0$ and $\kappa$ assuming values different from those originally considered (shortcutting a lot of computational work).  In doing so, we were able to show broadly, that the scaling of parameters (within an order of magnitude) could potentially save or slaughter a dubious result.  The effects from any viable cosmology~\cite{Clifton11,kostasrev} in these theories then becomes relevant, an example being applying Big Bang Nucleosynthesis (BBN) and Cosmic Microwave Background (CMB) constraints~\cite{Nconstraint}
to limit the effects of $G_N$ renormalisation - which we can convert into a bound on variations to $\kappa$.  Similarly constraints from fifth force experiments and bounds on anomalous accelerations on solar system scales could feed into our theories, but these would appear as parameters in any viable free function - something beyond the scope of our simple scaling argument.  

In this work, we aim to investigate the effects of varying the free functions in these theories (in what have become known as Type I theories of MOND, see~\cite{ali}).
%, although later we will make some more general statements on the validity
%of the results to other classes of MONDian theories).
 Here we show that a similar tidal stresses scaling algorithm exists in the linear regime of the theory, allowing us to rescale results previously computed
from a particular (fiducial) model of $\mu$ and in fact further symmetries present here can be exploited for computational gain also.  We also explore the behaviour of the (somewhat mysterious) deep MONDian scaling $C$, which has been determined numerically, but remains poorly constrained otherwise - here we consider how it gets rescaled between different models.  
 
 The structure of this paper is as follows: Firstly we consider analytical solutions in the linear regime of the theory for a simple free function parameterisation, showing how the functional forms of solutions vary in different regimes and
develop a new intermediate regime. In section \ref{transients} we consider an application of this work to transients for some more general $\mu$ and make considerations towards rescaling tidal stresses.  We also attempt to constrain the scalings in these models, comparing numerics in different models and providing some understanding as to their behaviour.  In the process, we develop a model independent framework for different free functions as a first step towards understanding the MONDian parameter space.  We finish with some broad conclusions and look towards future work.

%============================================================================
\section{Analytical Results}
In Type I MONDian theories, dynamics in the nonrelativistic regime result from the joint action of the usual Newtonian potential $\Phi_N$ (associated with the metric) and a ``fifth force'' scalar field, $\phi$. The total potential acting on nonrelativistic particles is their sum $\Phi=\Phi_N+\phi$.  The MONDian field $\phi$ is ruled by a nonlinear Poisson equation: \be \nabla \cdot \left(\mu(z)\nabla \phi\right) = \kappa G \rho, \ee with $z=\frac{\kappa}{4\pi}\frac{\vert\nabla\phi\vert}{a_0}$,
where $\kappa$ is a dimensionless constant and $a_0$ is the usual MONDian
acceleration.
We begin from the linear MONDian field \be \vU  = -\frac{\kappa}{4\pi}
\frac{\Del\phi}{a_0}\mu \label{U defn} \ee and will make use of the fact that since $U =
\mu z$, we can write any transition function also solely as $\mu = \mu(U)$.

This change of variable allows us to write dimensionless vacuum equations
as \bea \Del \cdot \vU &=& 0 \label{vect1} \\ 4\, m\, U^2 \Del \wedge \vU + \vU \wedge \Del U^2 &=& 0 \label{vect2}\eea with \be 4m = \frac{d \ln U^2}{d \ln \mu} \label{m eqn}\ee where $4m$ is picked for notational clarity later.

This system of equations is exactly that of the MONDian potential around the gravitational saddle points between two massive bodies - a truly low
potential test case and of considerable interest for a modified gravity test.  The background Newtonian force $F_N$ in such a system simply becomes linearised
 along the axes linking the two bodies and clearly must satisfy $\Del^2 \Phi_N = 0$.  Using spherical polar coordinates centered on the saddle, we find
\bea -\Del \Phi_N &=& \vF_N = A\, r\, \vN \\ \vN &=& N_r\ve_r + N_\psi \ve_\psi \\ N_r &=& \frac{1}{4}(1 + 3 \cos 2\psi ) \\ N_\psi &=& -\frac{3}{4} \sin 2\psi\eea where $A$ is just the tidal stress at the saddle and we notice due to the symmetries of this two body system, the polar angle $\varphi$ does not appear (although in general it would).  

Our next step is specify a $\mu$ function, consider the
different limits in such a system and find analytical solutions, as have been considered previously for one particular case of function~\cite{bekmag}, \be z = \frac{\mu}{\sqrt{1-\mu^4}} \Longleftrightarrow \mu = \sqrt{\frac{\sqrt{1+4z^4}-1}{2z^2}}\ee which we label $\mu_{fiducial}$ hereafter.  Here the aim is to look at parameterised functions, such as  \be \mu = \frac{z^n}{1 + z^n} \label{mu(n)} \Rightarrow 4m = \frac{2}{n}\left( n + \frac{1}{1 - \mu} \right) \ee and we will
consider further generalisations later.  Although we cannot (in
general) write a closed form for $m$ in terms of $U$, we find in the limits \bea z \ll 1 && \, \mu \approx z^n \Rightarrow 4m \approx \frac{2(n+1)}{n} \label{DM} \\ z \gg 1 && \,  \mu \approx 1 \Rightarrow 4m \approx \frac{2 \,U^n}{n}\label{QN} \eea Similarly the extra acceleration felt by test particles cannot generally be written down as a closed form expression in $U$, but
without loss of generality as \be \delta\vF = - \Del\phi = \frac{4 \pi a_0}{\kappa} \vU\left(1 + \frac{1}{z^n}\right) \label{deltaF} \ee and from equations (\ref{U defn}) and (\ref{mu(n)}) it is clear that \bea\nonumber z \gg 1 && z^n \approx U^n \\ \nonumber z \ll 1 && z^n \approx U^{n/(n+1)}\eea It is clear from $U = \mu z$ that each limit therefore satisfies \bea z \gg 1 &\Rightarrow& U \gg 1 \nonumber \\ z \ll 1 &\Rightarrow& U \ll 1 \nonumber\eea with the
obvious boundary between them located at \be |\vU|^2 \simeq 1 \Rightarrow r^2 \left(\cos^2 \psi + \frac{1}{4} \sin^2 \psi\right)^2 = \left(\frac{16\pi^2}{\kappa^2}\frac{a_0}{A}\right)^2
= r_0^2\ee which is just the equation for an ellipsoid with a semimajor axis
of size $r_0$.  These show that in general the functional forms of the inner
and outer ellipsoid solutions (hereafter bubble) should be quite different.  With these results in mind, let us proceed to finding the analytical solutions as before, using the linear Newtonian approximation.

\subsection{Quasi-Newtonian (QN) regime }\label{QN Analytic}
Given this system of vector equations, we need to specify boundary conditions.
 For $r/r_0 \gg 1$, we expect the MONDian potential to mimic the
 Newtonian $\phi \approx \frac{\kappa}{4\pi}\Phi_N$ and so our ansatz has to be of the form \bea \vU &=& \vU_0 + \vU_2 \\ \vU_0 &=& \frac{r}{r_0}\vN(\psi) \eea where $\vU_2$ will be some subdominant contribution as we move far from the saddle, but a very relevant one closer to the bubble.  Additionally although $\vU_0$ is curl free, the form of equation (\ref{vect2}) suggests $\vU_2$  could in general have a curl sourced by $\vU_0$, satisfying \bea \Del \cdot \vU_2 &=& 0 \label{divU2} \\ \Del \wedge \vU_2 &=& -\frac{\vU_0 \wedge \Del|\vU_0|^2}{2\,|\vU_0|^{n+2} /n } \label{curlU2} \eea Using the notation \be \vU_2 = U_r\ve_r + U_\psi\ve_\psi \ee equations (\ref{divU2}) and (\ref{curlU2}) take the form \bea
\frac{1}{r^2}\frac{\partial}{\partial r}(r^2 U_r)+\frac{1}{r\sin\psi}\frac{\partial}{\partial\psi}(\sin\psi\, U_\psi)&=&0
\label{div3}\\
\frac{1}{r}\left[\frac{\partial}{\partial r}(r U_\psi)-\frac{\partial U_r}{ \partial \psi}\right]&=& \frac{s_n(\psi)}{ r^n}
\label{curl3}
\eea
%with now generalised source term
\be
s_n(\psi)\equiv -3n \frac{2^{3 n/2}   \sin 2 \psi}{  (5 + 3 \cos 2 \psi)^{1 + n/2}}\label{source}
\ee and so we have to discuss the effect of varying $n$ on this extra curl
force.

\subsubsection{The $n=1$ case}                                           The source function here becomes \be s_1\equiv - \frac{6\sqrt{2} \sin 2\psi}{(5 + 3 \cos 2 \psi)^{3/2}}\label{sourcen1} \ee and equation (\ref{curl3}) suggests as ansatz of the form \be \vU_2 = \vB_1(\psi) = (F_1(\psi)\ve_r + G_1(\psi) \ve_\psi) \ee This reduces equation (\ref{curl3}) to \be G_1 = F_1' +  s_1\ee and hence equation (\ref{div3}) to \be 2F_1 + F_1' \cot\psi + F_1'' = -(s_1' + s_1 \cot\psi) \ee We can solve this using the standard techniques of inhomogeneous
ordinary differential equations (ODEs), to find expansions of $F$ and $G$ \bea F_1 &\approx& -0.23218 - 0.7201 \cos 2\psi + 0.1306 \cos 4\psi \nonumber \\ G_1 &\approx& 0.5115\sin 2\psi - 0.0556 \sin 4\psi  \eea 
%which we show in figure (\ref{fig:QNFG})
\begin{figure}
\begin{center}
\resizebox{1.\columnwidth}{!}{\includegraphics{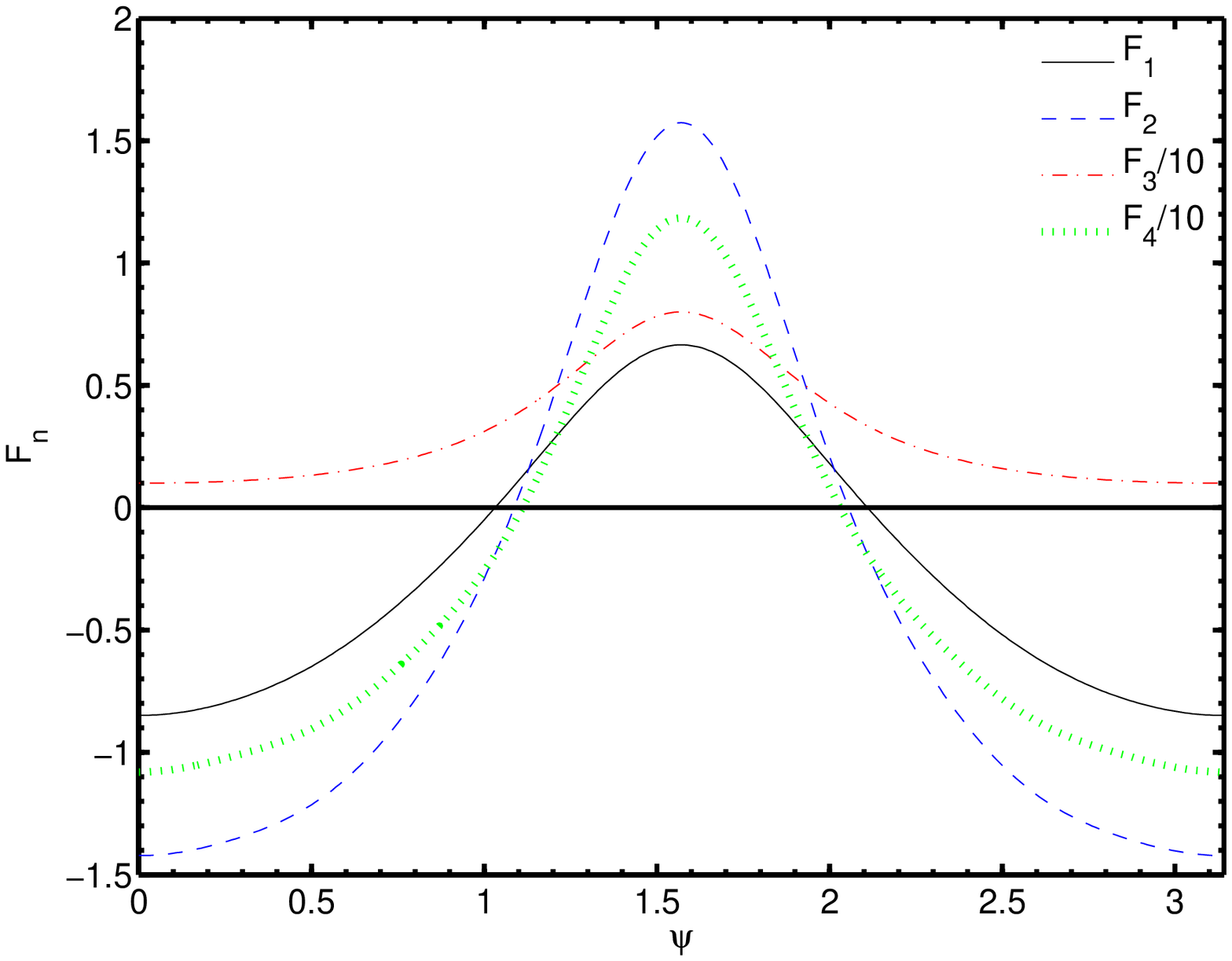}}
\resizebox{1.\columnwidth}{!}{\includegraphics{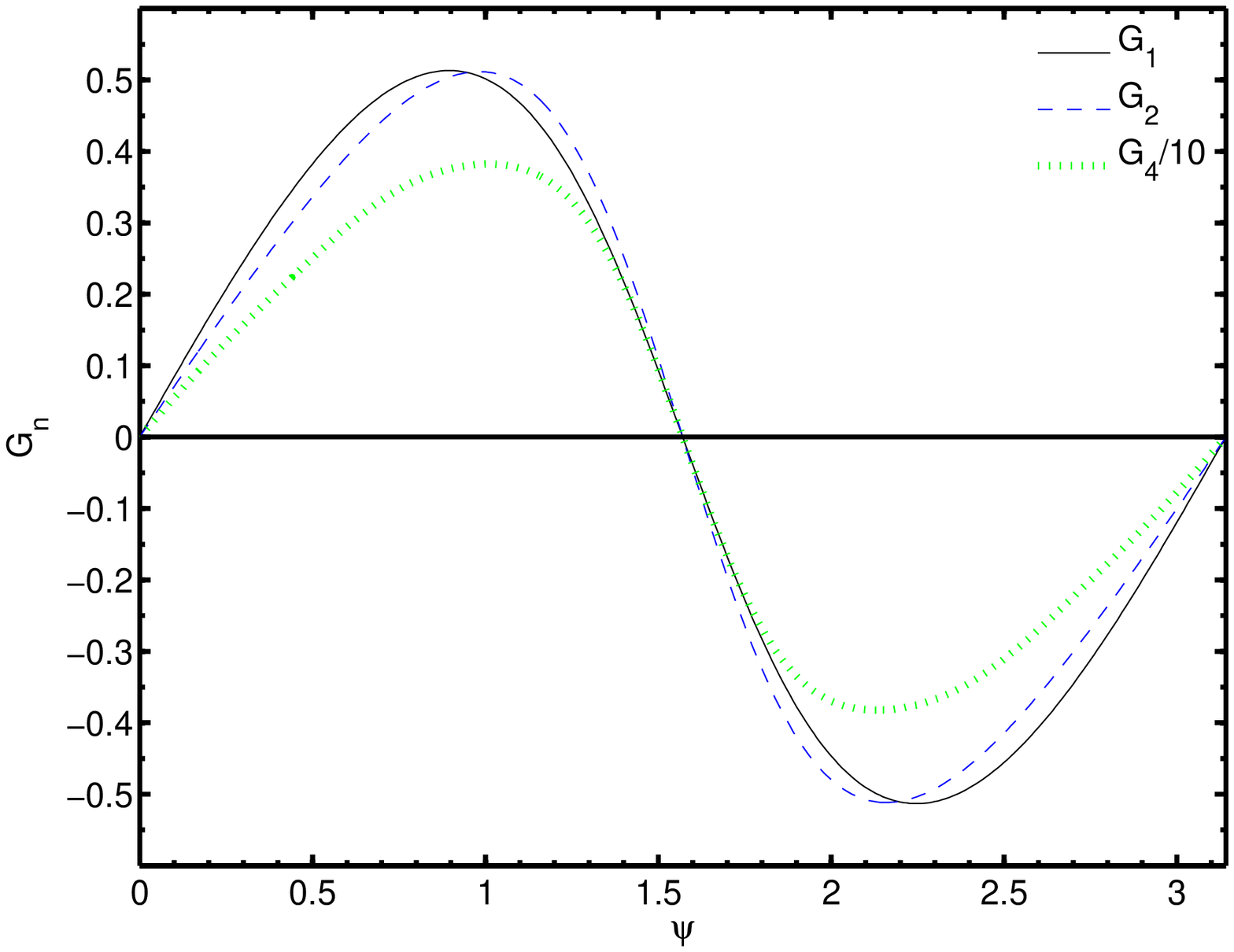}}
\caption{\label{fig:QNFG}{The angular profile functions F and G giving the direction of the curl field $B(\psi)$ in the QN region in a few different cases. We omit $G_3$ here due to it vanishing for all $\psi$. }}\end{center}
\end{figure}We find the extra acceleration felt by test particles is hence given by \be \delta\vF = - \Del\phi \approx \frac{4 \pi a_0}{\kappa} \vU\left(1 + \frac{1}{U}\right)\ee which for $U \gg 1$ suggests \be \delta F \approx \frac{4 \pi a_0}{\kappa} \left(\vU_0 + \frac{\vU_0}{U_0} + \vU_2 + ... \right) \label{deltaF QN}\ee
The first term, we call $\delta\vF_0$, is simply a fully Newtonian term which renormalises the gravitation constant.  
The second term, which we denote $\delta\vF_1$, is here just a rescaled unit vector of the Newtonian potential \be \delta\vF_1 = \frac{16 \pi a_0}{\kappa}\frac{\vN(\psi)}{\sqrt{10 + 6 \cos 2\psi}} \ee 
The third term, which we denote $\delta \vF_2$, is just the curl field contribution \be \delta \vF_2 = \frac{4 \pi a_0}{\kappa} \vB_1(\psi)\ee It is the terms $\delta \vF_1 + \delta \vF_2$ and higher which are the true MONDian observables in this system.  

\subsubsection{The $n=2$ case}
In this case, the source reduces to \be s_2\equiv -\frac{48\sin 2\psi}{ (5+3\cos2\psi)^2} \label{sourcen2} \ee which is just a rescaled solution of those from~\cite{bekmag}.
The form of equation (\ref{curl3}) here suggests both $U_r$ and $U_\psi$ behave as $1/r$, hence we can rewrite our ansatz as \be \vU_2 = \left(\frac{r_0}{r}\right) \vB_2(\psi) = \left(\frac{r_0}{r}\right)\left(F_2(\psi) \ve_r + G_2(\psi) \ve_\psi\right) \ee 
Summarising the results of the calculation here, we find the ansatz collapses equation (\ref{curl3}), allowing a simple separation of the components of $\vU$: \bea F_2 &=& \frac{8}{5 + 3 \cos 2\psi} + A \nonumber  \\ G_2\sin\psi&=&4\frac{\tan^{-1}(\sqrt 3 -2 \tan\frac{\psi}{2})+\tan^{-1}(\sqrt 3 +2 \tan\frac{\psi}{2})}{\sqrt 3}\nonumber\\ && + A\cos\psi + B \eea finding $A, B$ after imposing the conditions
of homogeneity and continuity and that at the boundaries of the bubbles we only have a radial force component (akin to the Newtonian), ie $G(\psi=0) = G(\psi=\pi)=0$,
\be \nonumber A = B = - \frac{4\pi}{3\sqrt 3}\ee 
We can then insert this expression into the expansion in equation (\ref{deltaF})
for $\delta \vF$ with $U \gg 1$ as before.

\subsubsection{The $n = 3$ case}
Here the source function becomes \be s_3 \equiv -\frac{144 \sqrt{2}\sin 2\psi}{(5 + 3\cos 2\psi )^{5/2}}\ee
We find solutions using a separable ansatz of the form \be \vU_2 = \left(\frac{r_0}{r}\right)^2
\vB_3(\psi) = \left(\frac{r_0}{r}\right)^2 (F_3(\psi) \ve_r + G_3(\psi) \ve_\psi) \ee with profile functions \bea F_3 &=& \frac{16\sqrt{2}}{(5 + 3\cos 2\psi)^{3/2}}
\nonumber \\ G_3 &=& 0\eea which satisfy the boundary conditions.

\subsubsection{Other cases}
For some power $n \geq 1$, we can make a more general ansatz as \be \vU_2 = \left(\frac{r_0}{r}\right)^{n-1} \vB_n(\psi) = 
\left(\frac{r_0}{r}\right)^{n-1} (F_n(\psi) \ve_r + G_n(\psi) \ve_\psi) \ee which we find now does not collapse equation (\ref{curl3}), but rather by combining with (\ref{div3}), we get a second order sourced ODE for $F_n$ in \be F_n (n-2)(n-3) + F_n' \cot \psi + F_n'' = -(s_n' + s_n\cot \psi) \label{ODEn} \ee and from (\ref{curl3}) \be (2-n)G_n - F_n' = s_n\ee The homogenous solutions of equation (\ref{ODEn}) are simply Legendre polynomials in $\cos \psi$ of order $(n-2)$ and the full inhomogenous solution can be found using standard ODE techniques.  
We find a generic feature of solutions in this regime is that \be (\delta \vF_1 + \delta \vF_2) \propto \frac{1}{r^{n-1}} \ee and since further terms in $\delta \vF$ tail off ever faster, the relative importance of the curl terms becomes diminished in the large $n$ limit.  Also our initial requirement
on $z$ that $n \geq 1$ means at worst $U_2 \propto r^0$, which will still
be washed out for $r/r_0 \gg 1$ against the Newtonian. 

\subsection{Deep MONDian (DM) regime}\label{DM Analytic}
Our previous intuition with boundary conditions does not help us here since
we expect a very different signal compared to the linear Newtonian falling
to zero at the saddle.  We can write equations (\ref{vect1}) and (\ref{vect2}) here as \bea \frac{1}{r^2}\frac{\partial}{\partial r}\left(r^2 \, U_r\right) + \frac{1}{r \sin \psi}\frac{\partial}{\partial \psi}(\sin \psi\, U_\psi) = 0 \,\,\,\,\,&&\\ \left[\frac{4m}{r}\left(\frac{\partial(r U_r)}{\partial r} - \frac{\partial U_\psi}{\partial \psi}\right) + \left(\frac{U_r}{r}\frac{\partial}{\partial \psi} - U_\psi \frac{\partial}{\partial r} \right) \right]U^2 = 0 \,\,\,\,\,&&\eea which given the scaling symmetries of these equations \bea \vU \rightarrow \vU \\ r \rightarrow \lambda\, r \eea suggests an ansatz for the potential as \be \vU = C \left(\frac{r}{r_0}\right)^{\alpha-2} (F(\psi) \ve_r + G(\psi) \ve_\psi) \ee where $\alpha - 2$ is used for notational convenience later
and $C$ is a constant required for matching between the two regimes.  We
can now use this ansatz to look for tidal stress solutions which keep $U$
small but become increasingly divergent as $r/r_0 \ll 1$.  Using this ansatz
gives a pair of coupled equations for $F$ and $G$ \bea  G' + G\, \cot(\psi) + \alpha F &=& 0 \label{alpha}\\ F \frac{d(F^2 + G^2)}{d \psi} + 2[\alpha' G - 2mF'](F^2 + G^2) &=& 0\label{alpha2} \eea where we are being clear to distinguish between the radial exponent $\alpha$ and the extended variable $\alpha'$ \be \alpha' = \alpha (2m-1) + 2(1-m) \label{alphaprime}\ee  Since we are seeking solutions where the tidal stresses diverge, we need to look at the additional force \be \delta \vF = -\Del \phi \approx \frac{4\pi a_0}{\kappa} \frac{\vU}{U^{\frac{n}{n+1}}} \label{gradphi DM}\ee which we rewrite in a
separable form as \be \delta \vF \approx \frac{4\pi a_0}{\kappa}\, C^{\frac{1}{n+1}}\left(\frac{r} {r_0}\right)^{\frac{\alpha-2}{n+1}} \frac{\vect{D}}{D^\frac{n}{n+1}}
\label{gradphi DM rD}\ee where $\vect{D}$ is the angular profile in the DM regime.  We see $\alpha < n + 3$ yields divergent tidal stress solutions. Requiring $n \geq 1$ puts bounds on $m$ as \be \frac{1}{2} < m \leq 1 \ee by substituting in $m$ and which after some manipulation, we find divergent solutions for
\be\alpha < \frac{6m-2}{2m-1} \ee and from inverting equation (\ref{alphaprime}) \be \alpha = \frac{\alpha' + 2(m-1)}{2m-1} \ee this makes our bound $\alpha' < 4m$ and hence given the bounds for $m$, we have that $\alpha' \leq 4$ is
always true. From similar considerations, we see that from requiring $U \ll 1$, $\alpha > 2$ in all cases (a point realised but not explicitly spelled out in~\cite{bekmag}).  These bounds are needed
in picking out the particular $\alpha$ we require from the sequence which satisfy the equations and permit regular solutions. 

% 
% We write the ansatz profile functions as fourier series and use equation (\ref{alpha}) to find algebraic relations between the fourier coefficients
% $f_i$ and $g_i$.  Equation (\ref{alpha2}) then yields quadratic and
% higher order equations in these coefficients, which are only solvable given
% certain values of $\alpha(n)$.  We use the boundary conditions, $F(\psi = 0) = F(\psi = \pi) = 1$ and $G(\psi = 0) = G(\psi = \pi) = 0$, to properly normalise each series.  

\subsubsection{The $m = 1$ case}
If we consider solutions with $m = 1$ (equivalent to $n=1$), then we are then guided to pick $\alpha' = \alpha \approx 3.528$ (hence the results from~\cite{bekmag} stand) with profile functions \bea F_1 &\approx& 0.2442 + 0.7246 \cos 2\psi + 0.0472 \cos 4\psi \nonumber \\ G_1 &\approx& -0.8334 \sin 2\psi - 0.0368 \sin 4\psi\eea

\subsubsection{The $\frac{1}{2} < m < 1$ case}
Here, we have $\alpha' \neq \alpha$ and so we need to find new solutions
to equations (\ref{alpha} - \ref{alpha2}).  We find solutions which neglect the derivative term in (\ref{alpha2})  as \bea F &=& a \cos \psi \nonumber \\ G &=& \mp a \sin \psi\eea (where $a$ is a constant) here with values of $\alpha_\pm$ given by \bea \alpha_+ &=& 2 \\  \alpha_- &=& \frac{2}{1-2m} \eea but in fact the $\alpha_-$ solution only exists when $m=1$, otherwise
equations (\ref{alpha} - \ref{alpha2}) are not be simultaneously satisfied.  We also find regular solutions exist for a discrete sequence of $\alpha(n)$'s
for each power $n$: $\{... , \alpha_{-1}, \alpha_\pm, \alpha_1, ...\}$, however
now we have lifted the degeneracy that $\alpha_{-i} = -\alpha_{i}$ (which
only exists in the $n=1$ case).  

\begin{figure}
\begin{center}
\resizebox{1.\columnwidth}{!}{\includegraphics{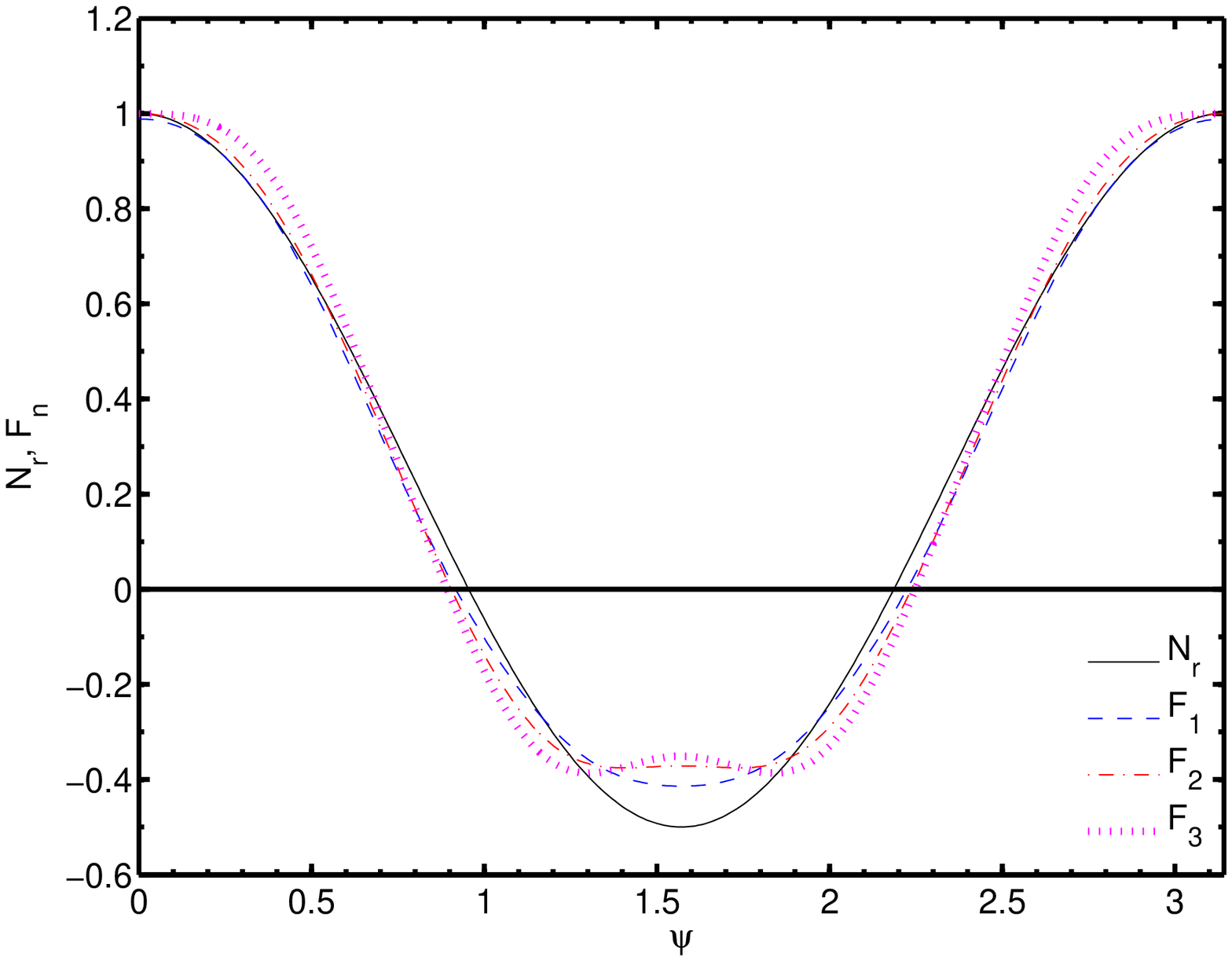}}
\resizebox{1.\columnwidth}{!}{\includegraphics{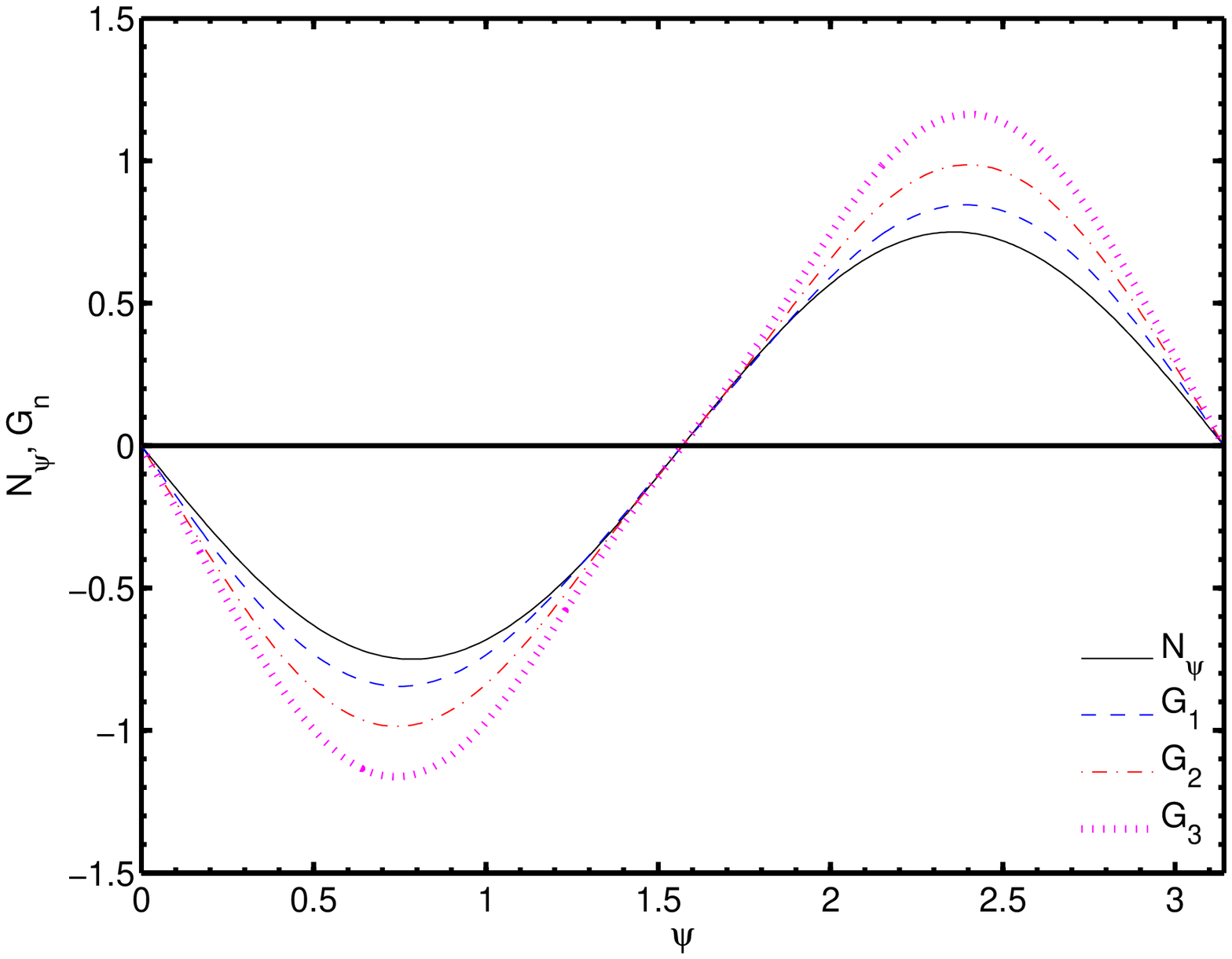}}
\caption{\label{fig:FnGn}{A few numerically determined angular profile functions
$F_n$ and $G_n$ in the DM region (dotted) compared with the Newtonian profile functions $N_r$ and $N_\psi$ (solid), in each figure respectively.
 Note the relative invariance of the radial profile and the slight changes in the azimuthal profile.}}\end{center}
\end{figure}

For $n = 2$, we have solutions for $\alpha_{-1} \approx -5.206$, $\alpha_1 \approx 3.983$ and we pick solutions where $2 < \alpha < 5$, nicely selecting out $\alpha_1$, with angular profile functions
as \bea F_2 &\approx&  0.248471 + 0.737261\cos 2\psi + 0.05982\cos 4 \psi \nonumber \\ G_2 &\approx& -0.9570 \sin 2\psi - 0.057766 \sin 4\psi \eea  
Similarly for $n=3$, we have $2 < \alpha < 6$ and so $\alpha_1 \approx 4.4057$, with angular profile functions \bea F_3 &\approx& 0.257381 + 0.766674 \cos 2\psi + 0.0624203 \cos 4\psi \nonumber \\ G_3 &\approx& -1.1099 \sin 2\psi - 0.0756068 \sin 4\psi  \eea
and we compare a few angular profiles in Figure \ref{fig:FnGn}.

%In each case, the divergence in the tidal stresses become stronger 

\subsubsection{The $m = \frac{1}{2}$ case} 
In this large $n$ limit, $\mu$ effectively becomes a step function.  We have
$\alpha'=1$ and find equation (\ref{alpha2}) reduces to \be F \frac{d(F^2 + G^2)}{d \psi} + 2[G - F'](F^2 + G^2) = 0 \ee which can be further manipulated to just \be \frac{d}{d\psi}\left(\frac{F}{G}\right) = 1 + \left(\frac{F}{G}\right)^2 \ee with the simple solution of \be F = G \tan (\psi + C_1) \ee and we pick $C_1 = \pm\pi/2$ to satisfy the boundary conditions.  Inserting this into equation (\ref{alpha}) gives solutions of the form \bea F &=& \mp C_2 \cos \psi\, (\sin \psi)^{\alpha-2} \nonumber \\ G &=& C_2 (\sin \psi)^{\alpha-1} \nonumber \eea This guides us to pick $C_2 = \mp 1$ and $\alpha = 2$ here, giving \bea
\nonumber F &=& \cos \psi \\ G &=& \mp \sin \psi \eea which mimic the $\alpha = 2$ solutions seen previously.  In this extreme case, the additional MONDian force is \be \delta \vF \rightarrow \frac{4\pi a_0}{\kappa} \frac{\vect{D}}{D}
\label{deltaF m=1/2}\ee which gives tidal stresses of the form \be S_{ij} \propto r^{-1} \ee which obviously diverges as we approach the saddle (although the divergence is now relatively weaker than the naive approach\footnote{The naive approach being to just take the `rule of thumb' modified inertia law, $F_\phi \propto \sqrt{F_N}$ for $a \leq a_0$, giving $S_{ij} \propto r^{-\epsilon}$ with $\epsilon = \frac{1}{2}$ here, however as we have shown, for $F_\phi^{n+1} \propto F_N$ with $n \geq 1$, we have solutions with $0.76 \lesssim \epsilon < 1$.} to MONDian tidal stresses suggests).

\subsubsection{Type II theories}

An important point to consider is what are the analogous effects of changing free functions in other classes of MONDian theories.  In the case of so-called Type II theories, changes to each of the regimes are easy to include in the computation of the sourced Poisson equation~\cite{aliqumond}.  Given here \bea \Del^2\phi = \frac{\kappa}{4\pi}\Del\cdot[\nu \Del\Phi_N] = r\,C_1\,\vN \cdot\Del \nu(r,\psi)  \eea where $\nu = \nu(v)$, $v = |\vU_0|$ and $C_1$ is some constant.  Solutions can be found with an ansatz \bea \phi = C_2\, r^a\, F_n(\psi) \eea where $C_2$ is another constant.  For some ``generalised'' DM limit of say $(F_\phi)^{n+1} \propto F_N$, \be\nu \rightarrow \left(\frac{1}{v}\right)^{\frac{n}{n+1}}\nonumber\ee
 This fixes the radial exponent $a$ as \be a = \frac{n+2}{n+1}\nonumber \ee and also fixes the sourced ODE for the profile function \be a(a+1)F_n + \cot\psi\, F_n' + F_n'' = g(\psi,n) \label{sourcedODE}\ee with generalised source term \be g(\psi,n) = (7+9\cos 2\psi) \left(\frac{2^{n-2}}{(5+3 \cos 2\psi)^{3n+2}}\right)^{\frac{1}{2(n+1)}}\ee
Given the results of sourced ODEs like this from section \ref{QN Analytic}, it seems unlikely therefore, that in this regime the solutions will remain similar, in stark contrast to the relative invariance of the DM limit in the Type I theories. In the QN regime, a similar system of ODEs exist, with varying source functions and parameters depending on the form of the falloff from $\nu \rightarrow 1$ and hence a similar conclusion can be reached.  A more detailed study of these generalised solutions however we leave for future work.

\subsection{An intermediate regime}
\begin{figure}[t]
\begin{center}
\resizebox{0.9\columnwidth}{!}{\includegraphics{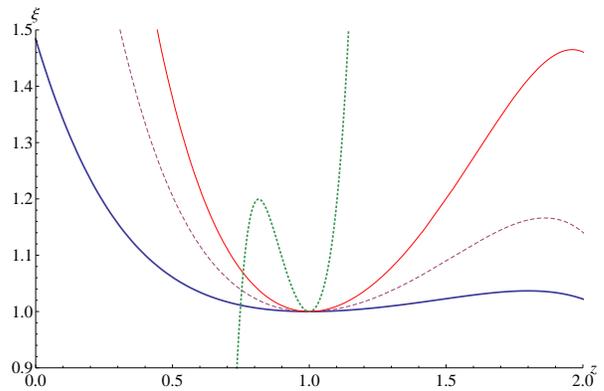}}
\caption{\label{fig:intermediate}{The ratio $\xi$ plotted around $z \sim
1$ for different values of $n$.  The blue (bold) line being $n=1$, the purple
(dashed) line $n=2$, the red (solid) line $n=3$ and the green (dotted) line $n=20$.  As we see, for small $n$, in this regime these models are relatively stable.}}\end{center}
\end{figure}

While we have a clear idea of the dynamics of $\vU$ in the large and small acceleration regimes, we lack much detail in the ``near field'' or intermediate regime (such as around $z \simeq 1$), except when we can estimate the size of the DM bubble.  Our work thus far has focused on finding the form and solutions to equation (\ref{vect2}) in each limit, however really we only need to start at the form of the $4m$ and see how it scales in each limit.  We can consider the leading order term to be of the form $\sim U^q$ (where $q$ is some power
to be found).  We begin with the expression \be \xi = \frac{4m}{C_q U^q}\ee where $C_q$ is simply some dimensionless constant and first derive our results in the two well understood regimes.  For the $\mu(n)$ models, \be \xi = \frac{2(n + 1 + z^n)}{n C_q z^q}\left(1 + \frac{1}{z^n}\right)^q\ee and we seek solutions
for $\xi \rightarrow 1$ in each limit.  In the $z \ll 1 $ regime, \be \xi \rightarrow \frac{2(n+1)}{n C_q}\frac{1}{z^{(n+1)q}}\ee which can only approach unity when $q = 0$ and \be C_q = \frac{2(n+1)}{n}\ee as before.  Similarly in the $ z \gg 1$ limit, \be \xi \rightarrow \frac{2}{n C_q} z^{n-q}\ee hence for unity $q = n$ and as before \be C_q = \frac{2}{n}\ee The real power of this technique can be exploited to attempt to solve these models
around some general point $z \simeq z_0$, which has the expansion: \bea \xi C_q|_{z_0} &\approx& 1 + \\\nonumber &+&  \frac{n z_0^n (1 + z_0^n) -  q (1 + n + z_0^n)^2}{z_0 (1 + z_0^n) (1 + n + z_0^n)}(z-z_0) \\ \nonumber &+& \mathcal{O}((z - z_0)^2)\eea with dimensionless scaling of the form
\be C_q|_{z_0} = \frac{2}{n} \left(\frac{z_0^{1+n}}{1+z_0^n}\right)^{-q}
(1 + z_0^n+n) \ee which we can solve at first order for $q$, \be q|_{z_0} = \frac{z_0^2(1+z_0^n) n}{(1+z_0^n+n)^2}\ee and seeing again that in the large and small $z_0$ limits, we recover the necessary behaviour for $U^q$. Around $z \simeq 1$, we find the behaviour \bea q &=& \frac{2 n}{(2 + n)^2} \\ C_q &=& \frac{(2 + n)}{n}2^{q+1}\eea and here $q$ is bounded for $n \geq 1$ as \be0 < q \leq \frac{2}{9}\nonumber\ee meaning that for large $n$, $C_q \rightarrow 2$, akin to an asymptotic DM regime in the large
$n$ limit.  We plot the full $\xi$ profile to demonstrate the relative stability of this limit
around $z\simeq1$ for different $n$'s in Figure \ref{fig:intermediate}. 

We see that, in general, $4m$ is not going to a constant here and so are guided to pick QN-like perturbative solutions, as before taking $\vU = \vU_0 + \vU_2$ with \be \vU_2 = \left(\frac{r}{r_0}\right)^{1-q} \vB_q(\psi)\ee

%============================================================================
\section{An Application - Transients}\label{transients}
One important use of this work is to look at how different power law falloffs and transients from $\phi \rightarrow \frac{\kappa}{4\pi}\Phi_N$ affect our results.  We can then examine how to rescale existing templates for
different $\mu$ functions.   Whilst the $\mu(n)$ model presents a nice parameterisation
here, we should also consider multiparameter families of functions, given that the various regimes are constrained by complementary but different
physical phenomena. 
Our `usual' DM limit, $\mu \propto z$, is motivated by the theory being a good dark matter replacement on low acceleration scales, a requirement we can drop {\it a priori}, work with some generalised limit, $\mu \propto z^n,$ and then introduce later (as necessary). While in the QN limit, the falloff from $\mu \rightarrow 1$ is governed by agreement with fifth force experiments in the solar system (in particular the strong requirement of no anomalous accelerations on scales set by the orbit of Neptune~\cite{ssconst}).  We
see, therefore, it is prudent to consider at least a two parameter family of free functions.  We can consider $\mu$'s of the form \be \mu = \frac{z^a}{(1+z^b)^{a/b}}\ee
% where this $\alpha$ must not be confused with the exponent $\alpha(n)$
% from the DM limit, $U_{DM} \propto r^{\alpha-2}$.  
and then as before, we compute  \bea 4m &=& \frac{2}{a}\left(a + \frac{1}{1-\mu^{b/a}} \right) \nonumber \\ \delta\vF &=& \frac{4 \pi a_0}{\kappa} \,\vU\left(1 + \frac{1}{z^b}\right)^{a/b}\eea where $z^b \approx U^b$ in the QN regime and  $z^b \approx U^{b/(a+1)}$ in the DM regime.  
In the case of these free functions, our intermediate scaling parameter
has the form around $z\simeq1$ of: \be q = \frac{2 b}{(2 + a)^2}\ee which is clearly greater than unity for $a = 1$ and $b \geq 5$, suggesting that the QN-like behaviour must have already been triggered before $z \simeq 1$, since these intermediate solutions already display the asymptotic-like behaviour (albeit with a ``stronger'' curl term - in the $z \gg 1$ regime here, $ U_2 \propto r^{1-b}$).

Our results are to be rescaled from our $\mu_{fiducial}$ model, presented in analytical and detailed numerical work previously~\cite{bekmag,bevis,ali}.
 In fact, if we pick some arbitrary model for $\mu(z)$, each regime takes the form \bea \mu \simeq \sum^\infty_{n=p} a_n z^n && z \ll 1 \label{DM series}\\ \mu \simeq 1 - \sum^\infty_{n=q} \frac{b_n}{z^n} && z \gg 1 \label{QN series}\eea and we consider each $a_n$ and $b_n$ as telling us the leading order ($n=p,q$) and higher terms of each expansion, with $p,q \geq 1$ in all cases.  In the DM regime, we find \be 4m \simeq 2\left(1 + \frac{\mu}{p\, z^{p}}\frac{1}{a_p}\right) \rightarrow \frac{2(p+1)}{p} \ee where $p$ is the exponent of the leading order term in $\mu(z)$.  We compute the term $\mu / z^{p}$ from equation (\ref{DM series}), however at leading order this is obviously a constant - explaining why there is little variability in this limit.  Similarly, in the QN limit, we find (using $U \simeq z$ here) \be 4m \simeq \frac{2}{U}\left(\frac{U^{q+1}}{q}\frac{1}{b_q}\right) \left(1 - \sum^\infty_{n=q+1} \frac{n b_n}{U^{n+1}} \right) \rightarrow \frac{2\, U^{q}}{q \,b_q} \label{genericQNmu}\ee where $q$ is the exponent of the leading order term in the expansion of $\mu(z)$.  As this result shows, here we are very much at the mercy of the free function we pick - naturally giving rise to the menagerie of solutions we found. 

Given these results, perhaps we can consider a similar scaling argument for
tidal stresses (as was considered in~\cite{MagAliscaling}).  A naive approach is to consider a window function (such as in section VC
of~\cite{ali}), ie one which preserves templates for $r < r_0$ (motivated
by the relative invariance of the profile functions in this limit) but rescales
them altogether differently outside of this.  There are however a couple of points that any rescaling algorithm needs to take into account: 
\begin{enumerate} \item{{\bf The Matching}} The constant $C$ in the DM solution is poorly constrained analytically and up to now has only been measured by looking at the ratio of numerical results to $C = 1$ analytical values
(over the range $r/r_0 =  0.05 \rightarrow 0.5$). Care needs to be taken that the correct normalisation for any rescaled tidal stresses is found. 

\item{{\bf Loss of Signal}} Since our analytical results are valid for $r/r_0
\ll 1$, the nature of the window function could cause us to loose some signal when we rescale.  Although these losses would be small compared to the signal deep inside the bubble, at the periphery and in the QN regime, noticeable
losses could occur if a naive rescaling is done.
\end{enumerate}
To investigate such issues, we need to consider how the tidal stresses change when we vary the azimuthal and radial components of the MONDian force ($\vect{F}_\phi(r,\psi)$).
%\subsection{Tidal Stresses}
Our anomalous tidal stresses are \be S_{ij} = -\frac{\partial^2\phi}{\partial
x_i \partial x_j} + \frac{\kappa}{4\pi}\frac{\partial^2 \Phi^N}{\partial x_i \partial x_j} \ee however for simplicity here we will compute just $S_{yy}(\vect{x})$,
although with a suitable coordinate change, any tidal stress component could be picked.  From the form of equation (\ref{gradphi DM rD}), we see we can write the MONDian force (in the linear regime) in the DM region as \bea F_r &=& C_1\, r^{\gamma}\, f(\psi) \nonumber \\ F_\psi &=& C_1\, r^\gamma\, g(\psi) \eea where \bea \gamma &=& \frac{\alpha(n) - 2}{n + 1}\nonumber \\\nonumber C_1 &=&  \frac{4 \pi a_0}{\kappa}\,\frac{C^{\frac{1}{n+1}}}{r_0^\gamma} =
\kappa A \frac{ C^{\frac{1}{n+1}} }{r_0^{\gamma-1}}\eea The tidal stresses %(subject to a change of coordinates) 
therefore are \bea S_{yy} &=& \frac{C_1}{2}r^{\gamma - 1} (f (1 + \gamma + (1 - \gamma) \cos 2\psi )  \\ &+& (g (\gamma - 1) + f') \sin 2\psi + 2 g' \cos^2\psi)) + \frac{C_2}{2} \nonumber\eea where $C_2 = \frac{\kappa}{4 \pi}A$ is the rescaled Newtonian tidal stress at the saddle.  We see clearly
the previous results~\cite{MagAliscaling} for changing $\kappa, a_0$ remain, ie $S_{yy} = \kappa A H_{yy}(r/r_0)$, but now also the effect of changing the exponent in $\mu \simeq z^n$ is clear.  As Figure \ref{fig:FnGn} shows, there is an approximate invariance of the profile functions in this regime, such that \bea F_n(\psi) &\simeq& F_1 \simeq N_r \nonumber \\ G_n(\psi) &\simeq& \xi(n) N_\psi \eea where $\xi(n)$ is some dimensionless linear scaling (for small $n$, we see $\xi\simeq1$).  As this shows, any naive window function rescaling of the tidal stresses runs the risk of ignoring relevant scalings dependent on $\gamma(\alpha(n))$ and $C$.  Similarly for the QN regime, in the linear regime (as for example
equation ({\ref{deltaF QN}) shows), the forces take the form \bea F_r &=& C_2 \,N_r(\psi)\, r + C_3 \,f(\psi)\,r^{1-n} \nonumber\\ F_\psi &=& C_2\, N_\psi(\psi)\, r + C_3\, g(\psi)\, r^{1-n} \eea with
\be C_3 = \frac{4 \pi a_0}{\kappa} r_0^{n-1} = \kappa A r_0^n\nonumber \ee and hence the tidal stresses are \bea S_{yy} &=& \frac{C_3}{2} r^{-n} (f(2 + n (\cos \psi - 1)) \\ \nonumber&+& 2 \cos \psi (\sin \psi (f' - n g) + g' \cos \psi )\eea 
 Here clearly we can try to play the same game with $f,g$ - however we are
 hampered by the fact that the curl term has more variance between models. % If however, we consider $n$ even, the scaling of $f$ and $g$ appears to
% follow \bea f_4 &\rightarrow& \chi_f\, f_2 \nonumber \\ g_4 &\rightarrow& %\chi_g\, g_2 \eea with $\chi_f \simeq 7.33$ and $\chi_g \simeq 7.50$.  In %general however, this appears not to be the case $\forall n$, as comparing %the profiles between $n=1,2$ models shows. 
%\subsection{The Deep MONDian Scaling}

Another issue that must be addressed is the effect of the MONDian scaling $C$ in these models.  It would be telling if the scaling could indeed be a function of the parameters (perhaps $n$) in some way.  Using the same techniques employed before, we reran our adaptive mesh code (presented in~\cite{bevis},
but with simple adaptations, see Appendix \ref{code}) with smaller grids, to compute the ratio of analytical results ({\it a priori} with $C=1$) and the numerical results for $r/r_0 = 0.1 \rightarrow 0.5$.  As Figure \ref{fig:C(n)} shows, the results are a little surprising.
\begin{figure}
\begin{center}
\resizebox{1.\columnwidth}{!}{\includegraphics{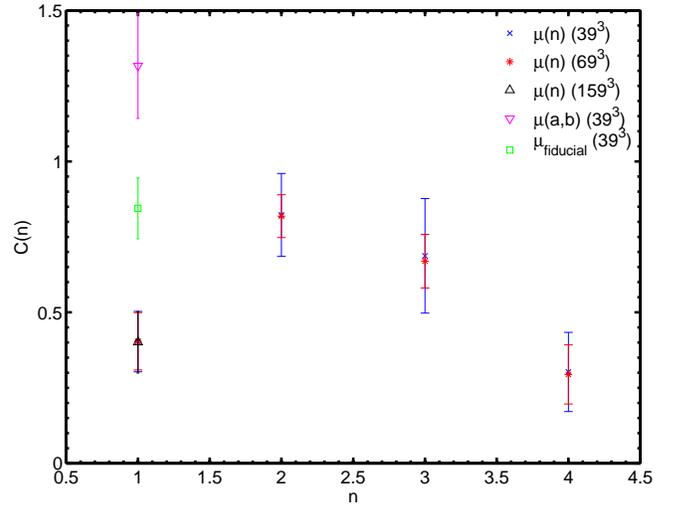}}
\caption{\label{fig:C(n)}{Deep MONDian scaling $C$ for different $\mu$ parameterisations, using different lattice sizes.  We measured the ratio of the linear solutions
compared to the full numerical solutions for $r/r_0 = 0.1 \rightarrow 0.5$
(since the smaller lattice sizes mean poorer resolution for small $r$, we
raised the starting cutoff).  As we see, whilst the errors reduce, the behaviour in $n$ remains as we increase the lattice size (suggesting these effects are not artifacts). For comparison, we also present the original scaling for our fiducial $\mu$, as well as a $\mu(a,b)$ model (with $a = 1$, $b = 2$)}.}\end{center}
\end{figure}  To try and make some sense of this, we need to look more closely at the matching between the regimes here.  We find the existence of three types of relevant constant in our system of equations:

\begin{itemize}
 \item {{ \bf DM regime}} For $z \ll 1$, $4m \rightarrow C^{DM}$, which becomes
 relevant when computing the exponent $\alpha(n)$ in the DM regime solutions.
   \item {{ \bf Departures from renormalised $G_N$}} \\
Expanding $\mu$ in the $z \gg 1 $ limit gives \be \mu^{-1} \simeq 1 + \frac{C_1^\mu}{z^p} +\frac{C_2^\mu}{z^{2p}}+ \dots\ee where for consistency we match each $C^\mu_i$
by expanding out $\mu^{-1}$ - we see that the coefficients parameterise the departures (at each order) from the usual Newtonian limit. 

  \item {{ \bf QN regime}} For $z \gg 1$, $4m \rightarrow U^p / C^{QN}$ where $p$ is the leading order power relevant in the expansion of $m$, such that when considering equation (\ref{curlU2}), \be \Del \wedge \vU_2 = \underbrace{- \frac{\vU_0 \wedge \Del|\vU_0|^2}{\,|\vU_0|^{p+2} }}_{\Large{\Del \wedge \vU_2^r}} C^{QN}\ee where $\vU_2^r$ is the (renormalised) curl term and $C^{QN}$ is the model dependent scaling.  As equation (\ref{genericQNmu}) shows however, this is related to $\mu$ as \be C^{QN} = \frac{p}{2}C^\mu_1\ee but we will
use this notation to be clear where each contribution arises from.  
\end{itemize}
Since we require the MONDian force be smooth and continuous, the matching between the different regimes must occur here.  Given \be \delta \vF = \frac{4\pi a_0}{\kappa} \frac{\vU}{\mu} \nonumber \ee put together with these model independent parameters, in the QN regime, reads as (the not unfamiliar expression of)

\bea \frac{\vU}{\mu} &\approx& \underbrace{\vU_0}_{\mathcal{O}(r^1)} + \underbrace{C_1^\mu \frac{\vU_0}{U_0^p} + C^{QN}\vU^r_2}_{\mathcal{O}(r^{1-p})} + \dots \eea where the higher order terms are $\mathcal{O}(r^{1-2p})$ or smaller and represent more complicated combinations of $U_0$ and $U_2^r$ (see Appendix \ref{QNDM
matching} for more details).  We matched this to the DM force \be \frac{\vU}{\mu} \approx C^{\frac{1}{\ell+1}}\,\left(\frac{r}{r_0}\right)^{\frac{\alpha-2}{\ell+1}} \frac{\vect{D_\ell}}{(D_\ell)^\frac{\ell}{\ell+1}} \nonumber \ee where for $z \ll 1$, $\mu = z^\ell + \dots$ and $D_\ell$ are just the angular
profiles for this case.  While matching the QN and DM regimes to find the scaling $C$ might seem dubious, note that we seek not to predict the actual values of $C(n)$ (since that would doubtless require knowledge of the system beyond linear order) but rather the scaling {\it between different} $C$'s.  In this way, we assume stricter validity in the linear regime but a more approximate one outside of this (as we have previously suggested with scaling rules). 
We summarise the results from such a matching and compare with values found
from numerical results: \begin{table}[H]
\caption{Comparison of Numerical and Matched $C$ Scalings} 
\centering  
\begin{tabular}{c c c c} 
%\hline  
$\mu$ & & $C_{num}$ & $C_{mat}$ \\ [0.5ex] \hline     
& & &\\
$\mu_{fiducial}$         &  & 0.8445 &  -\\[1ex] 
$\mu(n=1)$               &  & 0.4042 &  0.4050\\[1ex]
$\mu(n=2)$               &  & 0.8186 &  0.7903\\[1ex]
$\mu(a=1;b=2)$           &  & 1.3163 &  1.3723\\[1ex] 
\hline 
\end{tabular}
\label{table:C Scaling}
\end{table} and we provide more details about the matching in Appendix \ref{QNDM matching}, the broad conclusion being the variation in $C$ between models is an issue that can be dealt with.  As Table \ref{table:C Scaling} shows, we can predict relatively well how $C$ should scale without resorting to full numerical investigations.  
Naturally we should question that effectiveness of using these (so-called) ``shooting'' methods to interpolate between these disparate regimes.  Such techniques are commonplace in the field of numerical relativity, where matching asymptotic solutions to those close to a horizon, whilst maintaining regularity, is required. Recent work, however, has looked towards modeling curvature changes as akin to that of heat flows~\cite{wisemanRF1,wisemanRF2}.  Such Ricci flow techniques allow for a much cleaner determination of a systems dynamics, without the need for fine-tuning of parameters.  We leave the application of these spectral methods for future work. 

We can also consider free functions which do not asymptote to 1 (as explored in~\cite{Zhao, ali}).  We leave the full details for Appendix \ref{divergent mu}, the broad conclusions there being that instead of transitioning from a DM regime to a QN one, we move from one DM limit to a different one.  Additionally
the tidal stresses become suppressed by factors of $\left(\frac{\kappa}{4\pi}\right)^{1/2n}$ close to the saddle, meaning these models likely will not be well constrained by an LPF test (a conclusion previously reached by other methods in~\cite{ali}).

%============================================================================
\section{Constraining MOND}
A natural question to consider now is where do these results leave us when looking at data, such as measurements from LPF in the event of saddle flyby extension. {\it How (if at all) can we constrain our theories from data?  What exactly is the parameter space of MOND?}  
 
The most likely scenario for an LPF test is a single saddle flyby, which given its (likely large) velocity, suggests data collected with be on the time scale of minutes.  We posit that {\it if} any clear signal is seen above
the noise and Newtonian background, we can make the following inferences:

\begin{itemize}
\item{\bf The $0^{th}$ Order Approach}, for $b \lesssim r_0$, we assume that
the signal is dominated by the DM regime, which given the relative invariance of the profile functions (assuming $F_i(\psi) \simeq F_1$, $G_i(\psi) \simeq G_1$) means that the main scalings in the tidal stresses come from $\gamma$ and $C$.  

We can attempt to fit the DM tidal stresses to the signal by varying the value of $\gamma$ - remembering that really we have $\alpha(n)$ here and so really there only exists a strict series of $\gamma(\alpha(n))$ for regular solutions of $\vU$.  Once we find the correct $\gamma$, we can then consider the ``amplitude'' of this signal, which can tell us (albeit broadly) about the QN regime, from the matching between $C \rightarrow C^\mu_1$.

For $b \gtrsim r_0$, the signal is sampling the QN regime, which means at lowest order \be \frac{\vU}{\mu} - \vU_0 \simeq \frac{\vect{h}(\psi)}{(r/r_0)^{p-1}}\ee
where \be \vect{h}(\psi) = C^\mu_1 \left(\frac{\vN(\psi)}{N^p} + \frac{p}{2} \vB_p(\psi)\right) \ee Given that for small $n$, $h(\psi) \sim \mathcal{O}(1)$, we can first try fitting the radial falloff from the data and then once the exponent $p$ is found, the various angular profile functions can be inferred.  {\begin{center} \Large $\Downarrow$ \end{center}}

\item{\bf The $1^{st}$ Order Approach}, we introduce a cutoff in the signal, based on the impact parameter of the spacecraft (between the interior and exterior of the bubble).  We fit the DM signal as before (now taking into account the scaling of $G_i(\psi)$) and with the improved matching, make corrections to $C^\mu_1$.  

We consider each parameter can be written as a perturbative expansion as \be C^{\mu}_1 \simeq {C^{\mu}_1}^{\bf(0)} + {C^{\mu}_1}^{\bf(1)} + \dots \ee etc ... Allowing for improvements in parameter accuracy as we go up in approach. {\begin{center} \Large $\Downarrow$ \end{center}}

\item{\bf The $2^{nd}$ Order Approach}, if we are blessed with plenty of
tracking data and/or multiple flybys, we can make better determinations of our parameters.  
 
\end{itemize}

In this way, we can convert the distinctive MONDian signal of a positive result into a constraint on the MONDian parameter space akin to converting a negative result into a constraint on $\mu$ (as was considered in~\cite{ali}). %\newpage

%=========================================================================
\section{Conclusions}
To conclude, we have presented a series of techniques for characterising,
evaluating and rescaling the MONDian tidal stresses, which would be measured by LISA Pathfinder, should a saddle flyby be incorporated into the mission, for different models of $\mu$ function.   Our goal was to detach these theories from their ``alternative to dark matter duties'', considering two-regime functions with $\mu\rightarrow 1$ at large $z$, but $\mu\propto z^n$ (where in general $n\neq 1$), when $z$ is small.

Broadly speaking, we find an approximate invariance for the angular profile functions in the DM limit irrespective of the model of $\mu$ used, although changes do arise from the radial exponent $\gamma$ and DM scaling $C$.  In the QN limit, we find a menagerie of solutions depending on the form of the falloff from $\mu \rightarrow 1$. A very brief investigation into Type II MONDian theories suggests that the broad results here do not transfer over into that class of theory and so any rescaling of tidal stresses must be done more carefully, in contrast to the invariance we showed in Type I theories.   We also considered an intermediate MONDian limit, although it remains somewhat unclear what the solutions here can tell us about the transition between the other regimes.  

We suggested potential strategies to constrain the parameter space
of MONDian theories from data, identifying a framework of parameters that could be important for an experimental determination around saddle points.  In doing so, we demonstrated the interplay between the DM and QN limits by using this framework to calculate how the DM scaling $C$ varies in different models - a previously overlooked issue.  The framework also shows the possibility of calculating order by order the coefficient and falloff power in $\mu$ if {\it a priori} we assume values for $\kappa, a_0$ - at best we can constrain two parameters.  We leave the computation of scalings in SNRs for future work, as well as other applications of these techniques.

%=========================================================================
\begin{acknowledgments}
The author is grateful to Jo\~ao Magueijo for many fruitful suggestions and
advice, Alex Adam for useful discussions and Rocca di Papa for hospitality during the conception and preparation of this paper. The author is funded by an STFC studentship.  Some of the numerical work was carried out on the COSMOS supercomputer, which is supported by STFC, HEFCE and SGI.
\end{acknowledgments}

\appendix
%===============================================================================
\section{Adaptations to the Numerical Code}\label{code}

Minimal changes are required to run our code for these parameterised $\mu$ functions - the main difference being the term {\large$\frac{\partial\mu}{\partial g^2}$} in the computation of the discrete divergence on the lattice (see Appendix A of~\cite{bevis}),
for the $\mu(n)$ models:  \be \frac{\partial \mu}{\partial g^2} = \frac{n}{2}\left(\frac{\kappa}{4 \pi a_0}\right)^2 \mu^{1 - \frac{2}{n}} (1-\mu)^{1 + \frac{2}{n}} \ee and the $\mu(a, b)$ models: \be \frac{\partial \mu}{\partial g^2} = \frac{a}{2}\left(\frac{\kappa}{4 \pi a_0}\right)^2 \mu^{1 - \frac{2}{a}} (1-\mu^{\frac{b}{a}})^{1+\frac{2}{b}}\ee For completeness, we compare with the expression for $\mu_{fiducial}$: \be \frac{\partial \mu}{\partial g^2} = \frac{1}{2}\left(\frac{\kappa}{4 \pi a_0}\right)^2 \mu^{-1} (1-\mu^4)^{2} (1 + \mu^4)^{-1} \ee

%================================================================================
\section{Extending the QN $ \leftrightarrow $ DM Matching Calculation}
\label{QNDM matching}
Since we need to match the MONDian forces in each regime (rather than say $\vU$), we start from the expression\be \delta \vF \propto \frac{\vU}{\mu} \nonumber \ee along with \be \nonumber
\mu^{-1} \simeq 1 + \frac{C^\mu_1}{z^p} + \frac{C^\mu_2}{z^{2p}} + \dots \ee and given that \bea U &=& \sqrt{\vU\cdot\vU} \nonumber\\\nonumber  &=& \underbrace{|\vU_0|}_{\mathcal{O}(r^1)}\left(1 + 2\,C^{QN}\underbrace{\frac{\vU_0\cdot \vU_2^r}{U_0^2}}_{\mathcal{O}(r^{-p})}+ (C^{QN})^2\underbrace{\frac{\vU_2^r \cdot \vU_2^r}{U_0^2}}_{\mathcal{O}(r^{-2p})} \right)^{1/2} \\\eea we can put all of this together up to $3^{rd}$ order as

\bea \frac{\vU}{\mu} &\approx& \underbrace{\vU_0}_{\mathcal{O}(r^1)} + \underbrace{C_1^\mu \frac{\vU_0}{U_0^p} + C^{QN}\vU^r_2}_{\mathcal{O}(r^{1-p})} \\ &+& \underbrace{\frac{C_1^\mu C^{QN}}{U_0^p} \left(\vU_2^r - p \frac{\vU_0\cdot
\vU_2^r}{U_0^2}\vU_0\right) + C_2^\mu\frac{\vU_0}{U^{2p}_0}}_{\mathcal{O}(r^{1-2p})} \nonumber \\ &-& \underbrace{\frac{C^{QN}}{U_0^p}\left[\left(\frac{p}{2} C^\mu_1 \left(U_2^r\right)^2 - p_2\frac{(\vU_0\cdot\vU_2^r)^2}{U_0^2}\right)\frac{C^{QN}\vU_0}{U_0^2}\right.}_{\mathcal{O}(r^{1-3p})}
\nonumber \\ &+& \underbrace{\left. \vU_2^r\left(\frac{p}{2} C_1^\mu C^{QN}
\frac{\vU_0\cdot \vU_2^r}{U_0^{2}} - \frac{C^\mu_2}{U_0^{p}}\right) \right]}_{\mathcal{O}(r^{1-3p})} \nonumber \\ &+& \underbrace{\frac{\vU_0}{U_0^{2p}}\left(\frac{C^\mu_3}{U_0^{p}} - p C_2^\mu C^{QN} \frac{\vU_0\cdot \vU_2^r}{U_0^2} \right) }_{\mathcal{O}(r^{1-3p})} \,+ \,\dots  \nonumber \label{QN expansion C}\eea where $p_2 = p(p+2)/2$ and here the higher order terms are $\mathcal{O}(r^{1-4p})$ or smaller and represent much more complicated combinations of $U_0$ and $U_2^r$.  We summarise the values of these for various parameterisations in Table \ref{table:ConstantsParameters}.

\begin{table}[H]
\caption{Parameters for Various Models of $\mu$} 
\centering  
\begin{tabular}{c c c c} 
%\hline  
$C^\#$  & $\mu_{fiducial}$ & $\mu(n)$  & $\mu(a,b)$ \\ [0.5ex] \hline    & & &\\
$C^{DM}$        & 4                     & $\frac{2(n+1)}{n}$    & $\frac{2(a+1)}{a}$\\ [1ex] 
$C^{QN}$        & $\frac{1}{4}$         &  $\frac{n}{2}$        & $\frac{a}{2}$ \\[1ex] 
$C_1^{\mu}$       & $\frac{1}{4}$         & 1                     & $\frac{a}{b}$ \\ [1ex] 
$C_2^{\mu}$       & $\frac{1}{32}$         & 0                     & $\frac{a(a-b)}{2b^2}$ \\ [1ex] 
$C_3^{\mu}$       & $-\frac{1}{128}$         & 0                     & $\frac{a(a-b)(a-2b)}{3b^3}$ \\ [1ex] 
\hline 
\end{tabular}
\label{table:ConstantsParameters}
\end{table} \noindent and we bring attention to the fact that in the $\mu(a,b)$ model, the DM regime parameter $a$ plays a role in both regimes, whilst the $\mu(n)$ model is relatively constrained in the parameter space.  Using these parameters, we can perform the matching between the two regimes, here choosing the intermediate MONDian regime as \be\frac{\vU_{DM}}{\mu} \left[\frac{r}{r_0}\simeq 0.5\rightarrow 1\right]\underleftrightarrow{\;\;\; \mathbf{C^{\frac{1}{\ell+1}}}\;\; \;} \frac{\vU_{QN}}{\mu}\left[\frac{r}{r_0}\simeq 1 \leftarrow 3 \right]\nonumber \ee  and we present the results from such a matching and compare with numerical
values in Table \ref{table:Constants}.

\begin{table}[H]
\caption{Numerical and Matched $C$ Scalings} 
\centering  
\begin{tabular}{c c c c c c c} 
%\hline  
$\mu$ & & $C_{num}$ & $C_{mat}$ & ${\lambda}_{num}$ & ${\lambda}_{mat}$ &${C}^{improv}_{mat}$
\\ [0.5ex] \hline     
& & & & &\\
$\mu_{fiducial}$         &  & 0.8445 & 1.0523 & -       & -             &-\\[1ex] $\mu(n=1)$               &  & 0.4042 & 0.5046 & 2.0893  & 2.0852        &0.4050\\[1ex] $\mu(n=2)$               &  & 0.8186 & 0.9847 & 1.0316  & 1.0686        &0.7903\\[1ex]
$\mu(a=1;b=2)$           &  & 1.3163 & 1.7098 & 0.6416  & 0.6154        &1.3723\\[1ex] 
\hline 
\end{tabular}
\label{table:Constants}
\end{table} where $\lambda_\# = (C_{fiducial}/C_\mu)_\#$ represents the ratio between the fiducial DM scaling and the value for each model of $\mu$, in both the numerical and matched cases.  As we see the values of $C$ predicted in each case by matching typically overestimate its value.  The ratio between each value and the fiducial value (similarly obtaining by matching) is however much closer, suggesting we can find the scaling factor for the numerically
obtained $C_{fiducial}$ by, in general, computing how it scales in this matching process.

%============================================================================
\section{Divergent $\mu$ Models}\label{divergent mu}
We can, using these techniques, consider altogether different models of free function (such as those considered in~\cite{ali,Zhao}):
\be z = \frac{\kappa}{4\pi}\frac{\mu}{\frac{\kappa}{4\pi} + \beta \mu} \frac{1}{(1-\mu)^n} \ee Our central quantity for analysis has the form \be 4m = 2\left(2+\frac{\beta(n+1)\mu^2
+ (n\frac{\kappa}{4\pi}-\beta)\mu}{(1-\mu)(\frac{\kappa}{4\pi}+\beta \mu)} \right) \label{alpha 4m}\ee For $n=0$, these models display the asymptotic behaviour \bea \mu \ll 1 && \delta \vF \approx \left.\frac{4\pi a_0}{\kappa}\,
C^{\frac{1}{n+1}}\left(\frac{r} {r_0}\right)^{\frac{\alpha-2}{n+1}} \frac{\vect{D}}{D^\frac{n}{n+1}}\right|_{n=1} \\ \mu \gg 1 &&\delta \vF \approx \frac{a_0}{\beta}\frac{\vF_N}{F_N} 
\label{div mu>1} \eea with the feature that $z$ saturates as $F_N\rightarrow \infty$ (see~\cite{ali}, section VI A, in particular Fig 14), in this case we have \be 4m = 2\left(2 - \frac{\beta \mu}{\frac{\kappa}{4\pi} + \beta \mu}\right) = 2\left(2 - \frac{4\pi \beta}{\kappa} z\right) \ee in each limit becoming\bea \mu \ll 1 && 4m \simeq 4 \nonumber \\ \mu \gg 1 && 4m \simeq 2\nonumber \eea suggesting we are moving simply from one DM regime to a different one.  The form of the tidal
stresses for $\mu \gg 1$ are \bea S_{yy} &=& \frac{C_4}{r} \cos\psi
( (f+g')\cos \psi + \sin \psi \,(f' + g))  \nonumber \\&+& \frac{C_2}{2}\eea
where $C_4 = a_0 / \beta$ and as before $C_2$ is the rescaled Newtonian tidal
stress at the saddle.  We note that the magnitude of tidal stress scaling is suppressed by a factor of $\frac{\kappa}{4\pi}$ (compared with $C_1$).

For $n\neq0$, we find again the same DM limit $\mu \rightarrow z$, but now
a parameterised QN limit, of the form \bea \beta \gg \frac{\kappa}{4\pi}
&& \mu \simeq 1 - \left(\frac{\kappa}{4\pi\beta}\frac{1}{z} \right)^{1/n} + \dots  \nonumber \\ \beta \ll \frac{\kappa}{4\pi} && \mu \simeq 1 - \left(\frac{1}{z} \right)^{1/n} + \dots \eea in the QN limit, equation (\ref{alpha 4m}) reduces to \bea \beta \gg \frac{\kappa}{4\pi} &&4m \simeq 2n\left(\frac{4 \pi \beta}{\kappa}\right)^{1/n} U^{1/n} \nonumber \\ \beta \ll \frac{\kappa}{4\pi} && 4m \simeq 2n\, U^{1/n}
\\\nonumber \eea where we take notice of the limit relevant for the ``galactically preferred'' value of $\beta \approx 1$.  If we consider a simple case we have encountered before, say $n=1$, our model parameters now take the form \bea C^\mu_1 &=& \frac{\kappa}{4 \pi\beta} \ll 1\eea This strongly suggests that the DM scaling $C$ will be suppressed - allowing us to place some constraint on the combination of $\beta,n$ parameters here.   In this case we would expect \bea C &\rightarrow& \zeta C \\ \zeta
&\simeq& \left(\frac{\kappa}{4\pi}\right)^{2/n}\eea which in this example would be small and hence given this suppression in signal, such models could potentially evade the net of an LPF test.  For the case of $\beta \ll \frac{\kappa}{4\pi}$,
these functions simply fall into the cases we have described before (indeed our $\mu_{fiducial}$ follows a similar functional form).

%Too large an $\beta$ and we would need to fine tune $n$ depending on whether we see signal or not, however these results suggest a joint fit between galactic constraints (ie $\beta \approx  1$) and solar system constraints would be viable if LPF saw nothing.  

%==================================================================================
\bibliography{references}

\begin{thebibliography}{40}
\expandafter\ifx\csname natexlab\endcsname\relax\def\natexlab#1{#1}\fi
\expandafter\ifx\csname bibnamefont\endcsname\relax
  \def\bibnamefont#1{#1}\fi
\expandafter\ifx\csname bibfnamefont\endcsname\relax
  \def\bibfnamefont#1{#1}\fi
\expandafter\ifx\csname citenamefont\endcsname\relax
  \def\citenamefont#1{#1}\fi
\expandafter\ifx\csname url\endcsname\relax
  \def\url#1{\texttt{#1}}\fi
\expandafter\ifx\csname urlprefix\endcsname\relax\def\urlprefix{URL }\fi
\providecommand{\bibinfo}[2]{#2}
\providecommand{\eprint}[2][]{\url{#2}}

\bibitem[{\citenamefont{{D'Amico} et~al.}(2009)\citenamefont{{D'Amico},
  {Kamionkowski}, and {Sigurdson}}}]{DMastro}
\bibinfo{author}{\bibfnamefont{G.}~\bibnamefont{{D'Amico}}},
  \bibinfo{author}{\bibfnamefont{M.}~\bibnamefont{{Kamionkowski}}},
  \bibnamefont{and}
  \bibinfo{author}{\bibfnamefont{K.}~\bibnamefont{{Sigurdson}}}
  (\bibinfo{year}{2009}), \eprint{0907.1912}.

\bibitem[{\citenamefont{Milgrom}(1983)}]{Milgrom:1983ca}
\bibinfo{author}{\bibfnamefont{M.}~\bibnamefont{Milgrom}},
  \bibinfo{journal}{Astrophys. J.} \textbf{\bibinfo{volume}{270}},
  \bibinfo{pages}{365} (\bibinfo{year}{1983}).

\bibitem[{\citenamefont{Bekenstein and Milgrom}(1984)}]{aqual}
\bibinfo{author}{\bibfnamefont{J.}~\bibnamefont{Bekenstein}} \bibnamefont{and}
  \bibinfo{author}{\bibfnamefont{M.}~\bibnamefont{Milgrom}},
  \bibinfo{journal}{Astrophys. J.} \textbf{\bibinfo{volume}{286}},
  \bibinfo{pages}{7} (\bibinfo{year}{1984}).

\bibitem[{\citenamefont{Bekenstein}(2005)}]{teves}
\bibinfo{author}{\bibfnamefont{J.~D.} \bibnamefont{Bekenstein}},
  \bibinfo{journal}{{Phys. Rev. \textbf{D70}, 083509 (2004); Erratum-ibid.}}
  \textbf{\bibinfo{volume}{D71}}, \bibinfo{pages}{069901}
  (\bibinfo{year}{2005}), \eprint{astro-ph/0403694}.

\bibitem[{\citenamefont{{Jacobson} and {Mattingly}}(2001)}]{jacobmatAE}
\bibinfo{author}{\bibfnamefont{T.}~\bibnamefont{{Jacobson}}} \bibnamefont{and}
  \bibinfo{author}{\bibfnamefont{D.}~\bibnamefont{{Mattingly}}},
  \bibinfo{journal}{\prd} \textbf{\bibinfo{volume}{64}}, \bibinfo{eid}{024028}
  (\bibinfo{year}{2001}), \eprint{arXiv:gr-qc/0007031}.

\bibitem[{\citenamefont{Zlosnik et~al.}(2006)\citenamefont{Zlosnik, Ferreira,
  and Starkman}}]{aether}
\bibinfo{author}{\bibfnamefont{T.~G.} \bibnamefont{Zlosnik}},
  \bibinfo{author}{\bibfnamefont{P.~G.} \bibnamefont{Ferreira}},
  \bibnamefont{and} \bibinfo{author}{\bibfnamefont{G.~D.}
  \bibnamefont{Starkman}}, \bibinfo{journal}{Phys. Rev.}
  \textbf{\bibinfo{volume}{D74}}, \bibinfo{pages}{044037}
  (\bibinfo{year}{2006}), \eprint{gr-qc/0606039}.

\bibitem[{\citenamefont{Zlosnik et~al.}(2007)\citenamefont{Zlosnik, Ferreira,
  and Starkman}}]{aether1}
\bibinfo{author}{\bibfnamefont{T.~G.} \bibnamefont{Zlosnik}},
  \bibinfo{author}{\bibfnamefont{P.~G.} \bibnamefont{Ferreira}},
  \bibnamefont{and} \bibinfo{author}{\bibfnamefont{G.~D.}
  \bibnamefont{Starkman}}, \bibinfo{journal}{Phys. Rev.}
  \textbf{\bibinfo{volume}{D75}}, \bibinfo{pages}{044017}
  (\bibinfo{year}{2007}), \eprint{astro-ph/0607411}.

\bibitem[{\citenamefont{{Zuntz} et~al.}(2010)\citenamefont{{Zuntz}, {Zlosnik},
  {Bourliot}, {Ferreira}, and {Starkman}}}]{aether2}
\bibinfo{author}{\bibfnamefont{J.}~\bibnamefont{{Zuntz}}},
  \bibinfo{author}{\bibfnamefont{T.~G.} \bibnamefont{{Zlosnik}}},
  \bibinfo{author}{\bibfnamefont{F.}~\bibnamefont{{Bourliot}}},
  \bibinfo{author}{\bibfnamefont{P.~G.} \bibnamefont{{Ferreira}}},
  \bibnamefont{and} \bibinfo{author}{\bibfnamefont{G.~D.}
  \bibnamefont{{Starkman}}}, \bibinfo{journal}{\prd}
  \textbf{\bibinfo{volume}{81}}, \bibinfo{eid}{104015} (\bibinfo{year}{2010}),
  \eprint{1002.0849}.

\bibitem[{\citenamefont{{Bonvin} et~al.}(2008)\citenamefont{{Bonvin}, {Durrer},
  {Ferreira}, {Starkman}, and {Zlosnik}}}]{AESS}
\bibinfo{author}{\bibfnamefont{C.}~\bibnamefont{{Bonvin}}},
  \bibinfo{author}{\bibfnamefont{R.}~\bibnamefont{{Durrer}}},
  \bibinfo{author}{\bibfnamefont{P.~G.} \bibnamefont{{Ferreira}}},
  \bibinfo{author}{\bibfnamefont{G.}~\bibnamefont{{Starkman}}},
  \bibnamefont{and} \bibinfo{author}{\bibfnamefont{T.~G.}
  \bibnamefont{{Zlosnik}}}, \bibinfo{journal}{\prd}
  \textbf{\bibinfo{volume}{77}}, \bibinfo{eid}{024037} (\bibinfo{year}{2008}),
  \eprint{0707.3519}.

\bibitem[{\citenamefont{{Milgrom}}(2009)}]{bimetric}
\bibinfo{author}{\bibfnamefont{M.}~\bibnamefont{{Milgrom}}},
  \bibinfo{journal}{\prd} \textbf{\bibinfo{volume}{80}}, \bibinfo{eid}{123536}
  (\bibinfo{year}{2009}), \eprint{0912.0790}.

\bibitem[{\citenamefont{Sanders}(2005)}]{BSTV}
\bibinfo{author}{\bibfnamefont{R.~H.} \bibnamefont{Sanders}},
  \bibinfo{journal}{Mon. Not. Roy. Astron. Soc.}
  \textbf{\bibinfo{volume}{363}}, \bibinfo{pages}{459} (\bibinfo{year}{2005}),
  \eprint{astro-ph/0502222}.

\bibitem[{\citenamefont{Clifton et~al.}(2011)\citenamefont{Clifton, Ferreira,
  Padilla, and Skordis}}]{Clifton11}
\bibinfo{author}{\bibfnamefont{T.}~\bibnamefont{Clifton}},
  \bibinfo{author}{\bibfnamefont{P.~G.} \bibnamefont{Ferreira}},
  \bibinfo{author}{\bibfnamefont{A.}~\bibnamefont{Padilla}}, \bibnamefont{and}
  \bibinfo{author}{\bibfnamefont{C.}~\bibnamefont{Skordis}}
  (\bibinfo{year}{2011}), \eprint{1106.2476}.

\bibitem[{\citenamefont{{Famaey} and {McGaugh}}(2012)}]{Fam-gaugh}
\bibinfo{author}{\bibfnamefont{B.}~\bibnamefont{{Famaey}}} \bibnamefont{and}
  \bibinfo{author}{\bibfnamefont{S.~S.} \bibnamefont{{McGaugh}}},
  \bibinfo{journal}{Living Reviews in Relativity}
  \textbf{\bibinfo{volume}{15}}, \bibinfo{pages}{10} (\bibinfo{year}{2012}),
  \eprint{1112.3960}.

\bibitem[{\citenamefont{Zhao and Famaey}(2006)}]{Zhao}
\bibinfo{author}{\bibfnamefont{H.-S.} \bibnamefont{Zhao}} \bibnamefont{and}
  \bibinfo{author}{\bibfnamefont{B.}~\bibnamefont{Famaey}},
  \bibinfo{journal}{Astrophys. J.} \textbf{\bibinfo{volume}{638}},
  \bibinfo{pages}{L9} (\bibinfo{year}{2006}), \eprint{astro-ph/0512425}.

\bibitem[{\citenamefont{Famaey and Binney}(2005)}]{binney}
\bibinfo{author}{\bibfnamefont{B.}~\bibnamefont{Famaey}} \bibnamefont{and}
  \bibinfo{author}{\bibfnamefont{J.}~\bibnamefont{Binney}},
  \bibinfo{journal}{Mon. Not. Roy. Astron. Soc.}
  \textbf{\bibinfo{volume}{363}}, \bibinfo{pages}{603} (\bibinfo{year}{2005}),
  \eprint{astro-ph/0506723}.

\bibitem[{\citenamefont{Ferreras et~al.}(2008)\citenamefont{Ferreras,
  Sakellariadou, and Yusaf}}]{yusaf}
\bibinfo{author}{\bibfnamefont{I.}~\bibnamefont{Ferreras}},
  \bibinfo{author}{\bibfnamefont{M.}~\bibnamefont{Sakellariadou}},
  \bibnamefont{and} \bibinfo{author}{\bibfnamefont{M.~F.} \bibnamefont{Yusaf}},
  \bibinfo{journal}{Phys. Rev. Lett.} \textbf{\bibinfo{volume}{100}},
  \bibinfo{pages}{031302} (\bibinfo{year}{2008}), \eprint{0709.3189}.

\bibitem[{\citenamefont{Mavromatos et~al.}(2009)\citenamefont{Mavromatos,
  Sakellariadou, and Yusaf}}]{yusaf1}
\bibinfo{author}{\bibfnamefont{N.~E.} \bibnamefont{Mavromatos}},
  \bibinfo{author}{\bibfnamefont{M.}~\bibnamefont{Sakellariadou}},
  \bibnamefont{and} \bibinfo{author}{\bibfnamefont{M.~F.} \bibnamefont{Yusaf}},
  \bibinfo{journal}{Phys. Rev.} \textbf{\bibinfo{volume}{D79}},
  \bibinfo{pages}{081301} (\bibinfo{year}{2009}), \eprint{0901.3932}.

\bibitem[{\citenamefont{Ferreras et~al.}(2009)\citenamefont{Ferreras,
  Mavromatos, Sakellariadou, and Yusaf}}]{yusaf2}
\bibinfo{author}{\bibfnamefont{I.}~\bibnamefont{Ferreras}},
  \bibinfo{author}{\bibfnamefont{N.~E.} \bibnamefont{Mavromatos}},
  \bibinfo{author}{\bibfnamefont{M.}~\bibnamefont{Sakellariadou}},
  \bibnamefont{and} \bibinfo{author}{\bibfnamefont{M.~F.} \bibnamefont{Yusaf}},
  \bibinfo{journal}{Phys. Rev.} \textbf{\bibinfo{volume}{D80}},
  \bibinfo{pages}{103506} (\bibinfo{year}{2009}), \eprint{0907.1463}.

\bibitem[{\citenamefont{Angus et~al.}(2006)\citenamefont{Angus, Famaey, and
  Zhao}}]{Angus}
\bibinfo{author}{\bibfnamefont{G.~W.} \bibnamefont{Angus}},
  \bibinfo{author}{\bibfnamefont{B.}~\bibnamefont{Famaey}}, \bibnamefont{and}
  \bibinfo{author}{\bibfnamefont{H.}~\bibnamefont{Zhao}},
  \bibinfo{journal}{Mon. Not. Roy. Astron. Soc.}
  \textbf{\bibinfo{volume}{371}}, \bibinfo{pages}{138} (\bibinfo{year}{2006}),
  \eprint{astro-ph/0606216}.

\bibitem[{\citenamefont{Clowe et~al.}(2006)}]{bullet}
\bibinfo{author}{\bibfnamefont{D.}~\bibnamefont{Clowe}} \bibnamefont{et~al.},
  \bibinfo{journal}{Astrophys. J.} \textbf{\bibinfo{volume}{648}},
  \bibinfo{pages}{L109} (\bibinfo{year}{2006}), \eprint{astro-ph/0608407}.

\bibitem[{\citenamefont{Dai et~al.}(2008)\citenamefont{Dai, Matsuo, and
  Starkman}}]{bullet1}
\bibinfo{author}{\bibfnamefont{D.-C.} \bibnamefont{Dai}},
  \bibinfo{author}{\bibfnamefont{R.}~\bibnamefont{Matsuo}}, \bibnamefont{and}
  \bibinfo{author}{\bibfnamefont{G.}~\bibnamefont{Starkman}},
  \bibinfo{journal}{Phys. Rev.} \textbf{\bibinfo{volume}{D78}},
  \bibinfo{pages}{104004} (\bibinfo{year}{2008}), \eprint{0806.4319}.

\bibitem[{\citenamefont{Angus and McGaugh}(2007)}]{bullet2}
\bibinfo{author}{\bibfnamefont{G.~W.} \bibnamefont{Angus}} \bibnamefont{and}
  \bibinfo{author}{\bibfnamefont{S.~S.} \bibnamefont{McGaugh}}
  (\bibinfo{year}{2007}), \eprint{0704.0381}.

\bibitem[{\citenamefont{Brownstein and Moffat}(2007)}]{bullet3}
\bibinfo{author}{\bibfnamefont{J.~R.} \bibnamefont{Brownstein}}
  \bibnamefont{and} \bibinfo{author}{\bibfnamefont{J.~W.}
  \bibnamefont{Moffat}}, \bibinfo{journal}{Mon. Not. Roy. Astron. Soc.}
  \textbf{\bibinfo{volume}{382}}, \bibinfo{pages}{29} (\bibinfo{year}{2007}),
  \eprint{astro-ph/0702146}.

\bibitem[{\citenamefont{{Withers}}(2009)}]{withers}
\bibinfo{author}{\bibfnamefont{B.}~\bibnamefont{{Withers}}},
  \bibinfo{journal}{Classical and Quantum Gravity}
  \textbf{\bibinfo{volume}{26}}, \bibinfo{pages}{225009}
  (\bibinfo{year}{2009}), \eprint{0905.2446}.

\bibitem[{\citenamefont{{Zuntz} et~al.}(2008)\citenamefont{{Zuntz}, {Ferreira},
  and {Zlosnik}}}]{lviolationcosmo}
\bibinfo{author}{\bibfnamefont{J.~A.} \bibnamefont{{Zuntz}}},
  \bibinfo{author}{\bibfnamefont{P.~G.} \bibnamefont{{Ferreira}}},
  \bibnamefont{and} \bibinfo{author}{\bibfnamefont{T.~G.}
  \bibnamefont{{Zlosnik}}}, \bibinfo{journal}{Physical Review Letters}
  \textbf{\bibinfo{volume}{101}}, \bibinfo{eid}{261102} (\bibinfo{year}{2008}),
  \eprint{0808.1824}.

\bibitem[{\citenamefont{Blanchet and Novak}(2011)}]{Blanchet}
\bibinfo{author}{\bibfnamefont{L.}~\bibnamefont{Blanchet}} \bibnamefont{and}
  \bibinfo{author}{\bibfnamefont{J.}~\bibnamefont{Novak}}
  (\bibinfo{year}{2011}), \eprint{1105.5815}.

\bibitem[{\citenamefont{Sereno and Jetzer}(2006)}]{Sereno}
\bibinfo{author}{\bibfnamefont{M.}~\bibnamefont{Sereno}} \bibnamefont{and}
  \bibinfo{author}{\bibfnamefont{P.}~\bibnamefont{Jetzer}},
  \bibinfo{journal}{Mon. Not. Roy. Astron. Soc.}
  \textbf{\bibinfo{volume}{371}}, \bibinfo{pages}{626} (\bibinfo{year}{2006}),
  \eprint{astro-ph/0606197}.

\bibitem[{\citenamefont{Milgrom}(2009)}]{Milgromss}
\bibinfo{author}{\bibfnamefont{M.}~\bibnamefont{Milgrom}}
  (\bibinfo{year}{2009}), \eprint{0906.4817}.

\bibitem[{\citenamefont{McNamara et~al.}(2008)\citenamefont{McNamara, Vitale,
  and Danzmann}}]{LISA}
\bibinfo{author}{\bibfnamefont{P.}~\bibnamefont{McNamara}},
  \bibinfo{author}{\bibfnamefont{S.}~\bibnamefont{Vitale}}, \bibnamefont{and}
  \bibinfo{author}{\bibfnamefont{K.}~\bibnamefont{Danzmann}}
  (\bibinfo{collaboration}{LISA}), \bibinfo{journal}{Class. Quant. Grav.}
  \textbf{\bibinfo{volume}{25}}, \bibinfo{pages}{114034}
  (\bibinfo{year}{2008}).

\bibitem[{\citenamefont{Trenkel et~al.}(2009)\citenamefont{Trenkel, Kemble,
  Bevis, and Magueijo}}]{companion}
\bibinfo{author}{\bibfnamefont{C.}~\bibnamefont{Trenkel}},
  \bibinfo{author}{\bibfnamefont{S.}~\bibnamefont{Kemble}},
  \bibinfo{author}{\bibfnamefont{N.}~\bibnamefont{Bevis}}, \bibnamefont{and}
  \bibinfo{author}{\bibfnamefont{J.}~\bibnamefont{Magueijo}},
  \bibinfo{journal}{submitted}  (\bibinfo{year}{2009}).

\bibitem[{\citenamefont{Bevis et~al.}(2010)\citenamefont{Bevis, Magueijo,
  Trenkel, and Kemble}}]{bevis}
\bibinfo{author}{\bibfnamefont{N.}~\bibnamefont{Bevis}},
  \bibinfo{author}{\bibfnamefont{J.}~\bibnamefont{Magueijo}},
  \bibinfo{author}{\bibfnamefont{C.}~\bibnamefont{Trenkel}}, \bibnamefont{and}
  \bibinfo{author}{\bibfnamefont{S.}~\bibnamefont{Kemble}},
  \bibinfo{journal}{Class. Quant. Grav.} \textbf{\bibinfo{volume}{27}},
  \bibinfo{pages}{215014} (\bibinfo{year}{2010}), \eprint{0912.0710}.

\bibitem[{\citenamefont{Bekenstein and Magueijo}(2006)}]{bekmag}
\bibinfo{author}{\bibfnamefont{J.}~\bibnamefont{Bekenstein}} \bibnamefont{and}
  \bibinfo{author}{\bibfnamefont{J.}~\bibnamefont{Magueijo}},
  \bibinfo{journal}{Phys. Rev.} \textbf{\bibinfo{volume}{D73}},
  \bibinfo{pages}{103513} (\bibinfo{year}{2006}), \eprint{astro-ph/0602266}.

\bibitem[{\citenamefont{Magueijo and Mozaffari}(2012)}]{ali}
\bibinfo{author}{\bibfnamefont{J.}~\bibnamefont{Magueijo}} \bibnamefont{and}
  \bibinfo{author}{\bibfnamefont{A.}~\bibnamefont{Mozaffari}},
  \bibinfo{journal}{Phys. Rev. D} \textbf{\bibinfo{volume}{85}},
  \bibinfo{pages}{043527} (\bibinfo{year}{2012}), \eprint{1107.1075}.

\bibitem[{\citenamefont{{Magueijo} and {Mozaffari}}(2012)}]{MagAliscaling}
\bibinfo{author}{\bibfnamefont{J.}~\bibnamefont{{Magueijo}}} \bibnamefont{and}
  \bibinfo{author}{\bibfnamefont{A.}~\bibnamefont{{Mozaffari}}},
  \bibinfo{journal}{Phys. Rev. D} \textbf{\bibinfo{volume}{86}},
  \bibinfo{pages}{123518} (\bibinfo{year}{2012}), \eprint{1204.6663}.

\bibitem[{\citenamefont{Skordis}(2009)}]{kostasrev}
\bibinfo{author}{\bibfnamefont{C.}~\bibnamefont{Skordis}},
  \bibinfo{journal}{Class. Quant. Grav.} \textbf{\bibinfo{volume}{26}},
  \bibinfo{pages}{143001} (\bibinfo{year}{2009}), \eprint{0903.3602}.

\bibitem[{\citenamefont{Umezu et~al.}(2005)\citenamefont{Umezu, Ichiki, and
  Yahiro}}]{Nconstraint}
\bibinfo{author}{\bibfnamefont{K.-i.} \bibnamefont{Umezu}},
  \bibinfo{author}{\bibfnamefont{K.}~\bibnamefont{Ichiki}}, \bibnamefont{and}
  \bibinfo{author}{\bibfnamefont{M.}~\bibnamefont{Yahiro}},
  \bibinfo{journal}{Phys. Rev. D} \textbf{\bibinfo{volume}{72}},
  \bibinfo{pages}{044010} (\bibinfo{year}{2005}), \eprint{astro-ph/0503578v1}.

\bibitem[{\citenamefont{Mozaffari}(2011)}]{aliqumond}
\bibinfo{author}{\bibfnamefont{A.}~\bibnamefont{Mozaffari}},
  \bibinfo{journal}{arXiv}  (\bibinfo{year}{2011}), \eprint{1112.5443}.

\bibitem[{\citenamefont{Sanders}(2006)}]{ssconst}
\bibinfo{author}{\bibfnamefont{R.}~\bibnamefont{Sanders}},
  \bibinfo{journal}{Mon.Not.Roy.Astron.Soc.} \textbf{\bibinfo{volume}{370}},
  \bibinfo{pages}{1519} (\bibinfo{year}{2006}), \eprint{astro-ph/0602161}.

\bibitem[{\citenamefont{{Headrick} et~al.}(2010)\citenamefont{{Headrick},
  {Kitchen}, and {Wiseman}}}]{wisemanRF1}
\bibinfo{author}{\bibfnamefont{M.}~\bibnamefont{{Headrick}}},
  \bibinfo{author}{\bibfnamefont{S.}~\bibnamefont{{Kitchen}}},
  \bibnamefont{and}
  \bibinfo{author}{\bibfnamefont{T.}~\bibnamefont{{Wiseman}}},
  \bibinfo{journal}{Classical and Quantum Gravity}
  \textbf{\bibinfo{volume}{27}}, \bibinfo{pages}{035002}
  (\bibinfo{year}{2010}), \eprint{0905.1822}.

\bibitem[{\citenamefont{{Adam} et~al.}(2012)\citenamefont{{Adam}, {Kitchen},
  and {Wiseman}}}]{wisemanRF2}
\bibinfo{author}{\bibfnamefont{A.}~\bibnamefont{{Adam}}},
  \bibinfo{author}{\bibfnamefont{S.}~\bibnamefont{{Kitchen}}},
  \bibnamefont{and}
  \bibinfo{author}{\bibfnamefont{T.}~\bibnamefont{{Wiseman}}},
  \bibinfo{journal}{Classical and Quantum Gravity}
  \textbf{\bibinfo{volume}{29}}, \bibinfo{pages}{165002}
  (\bibinfo{year}{2012}), \eprint{1105.6347}.

\end{thebibliography}

\end{document}